\documentclass[aps, prd, floats, floatfix, twocolumn, superscriptaddress, nofootinbib, showpacs, letterpaper,10pt]{revtex4-1}

\usepackage{graphicx}
\usepackage{amsmath}
\usepackage{amsfonts}
\usepackage{amssymb}
\usepackage{amstext}
\usepackage{tensor}
\usepackage[margin=0.95in]{geometry}
\usepackage[english]{babel}
\usepackage{blindtext}
\usepackage{helvet}
\usepackage{microtype}
\usepackage[pdftex,
  pdftitle={Cosmic String Loop Microlensing},
  pdfauthor={Jolyon K. Bloomfield and David F. Chernoff},
  bookmarks,bookmarksopen=false,
  pdfstartview={FitH},
  linktocpage=true
]{hyperref}

\newcommand{\bi}{\begin{itemize}}
\newcommand{\ei}{\end{itemize}}
\def\be{\begin{align}}
\def\ee{\end{align}}
\def\bea#1\eea{\begin{align}#1\end{align}}
\def\bse{\begin{subequations}}
\def\ese{\end{subequations}}
\newcommand{\barr}{\begin{array}}
\newcommand{\earr}{\end{array}}

\newcommand{\simless}[0]{\mathbin{\lower 3pt\hbox
   {$\rlap{\raise 5pt\hbox{$\char'074$}}\mathchar"7218$}}}
\newcommand{\simgreat}[0]{\mathbin{\lower 3pt\hbox
   {$\rlap{\raise 5pt\hbox{$\char'076$}}\mathchar"7218$}}}

\begin{document}

%\title{Numerical Realizations of Cosmic String Microlensing}
\title{Cosmic String Loop Microlensing}

\author{Jolyon K. Bloomfield}
\affiliation{MIT Kavli Institute for Astrophysics and Space Research, Massachusetts Institute of Technology, Cambridge, MA 02139, USA}
\email{jolyon@mit.edu}

\author{David F. Chernoff}
\affiliation{Center for Radiophysics and Space Research, Cornell University, Ithaca, NY 14853, USA}
\email{chernoff@astro.cornell.edu}

\date{\today}

\begin{abstract}
Cosmic superstring loops within the galaxy
microlens background point sources lying close to the
observer-string line of sight. For suitable alignments, multiple paths coexist and the
(achromatic) flux enhancement is a factor of two. We explore this unique type of lensing by
numerically solving for geodesics that extend from source to
observer as they pass near an oscillating string. We
characterize the duration of the flux doubling and the scale
of the image splitting. We probe and confirm the existence
of a variety of fundamental effects predicted from previous
analyses of the static infinite straight string: the deficit
angle, the Kaiser-Stebbins effect, and the scale of the
impact parameter required to produce microlensing. Our
quantitative results for dynamical loops vary by $O(1)$
factors with respect to estimates based on infinite straight strings for a given impact parameter. A number of new features are identified in the
computed microlensing solutions. Our results suggest that
optical microlensing can offer a new and potentially
powerful methodology for searches for superstring loop
relics of the inflationary era.
\end{abstract}

% 04.25.D-: Numerical relativity
% 98.80.Cq: Particle-theory and field-theory models of the early Universe (including cosmic pancakes, cosmic strings, chaotic phenomena, inflationary universe, etc.)

\maketitle

\section{Introduction}
Shortly after its birth, the universe is believed to have
grown exponentially in size via an inflationary mechanism.
The almost scale-invariant density perturbation spectrum
predicted by inflation is strongly supported by observations of the
cosmic microwave background radiation carried out by the WMAP \cite{WMAP} and Planck satellites \cite{Ade:2013uln}. String theory, the best-developed tool to explore this
epoch, suggests the birth and survival of a network of
one-dimensional structures on cosmological scales \cite{Copeland:2003bj}.
Numerous fossil remnants of these cosmic superstrings may exist
within the galaxy, which could be revealed through the optical
lensing of background stars by way of the technique of
microlensing \cite{Chernoff2007}.
This paper investigates the fundamental physical features of
microlensing relevant to optical searches for string loops
generated by the network.

The primary parameter that controls string network and cosmic loop properties and dynamics is the string tension, $\mu$. The first exploration of strings was in the context of phase transitions in grand unified field theories (GUTs), which generated horizon-crossing defects with tension $G \mu/c^2 \sim 10^{-6}$ (hereafter, $c=1$) set by the characteristic grand unification energy \cite{Vilenkin1994}. Such defects could seed the density of fluctuations for galaxies and clusters, and would also create cosmic string loops.

The lifetime of a loop decaying by gravitational evaporation scales as $\mu^{-1}$. With large string tensions, GUT strings decay quickly, and do not live many Hubble expansion times. This implies that for such loops, clustering is largely irrelevant: they move rapidly at birth, become briefly damped by cosmic expansion, and are re-accelerated to relativistic velocities by the momentum recoil of anisotropic gravitational wave emission (known as the `rocket effect') before fully evaporating. As such, GUT loops were thought to be homogeneously distributed throughout space \cite{Vachaspati1985}.

However, empirical upper bounds on $\mu$ from a number of experiments in the past decade have essentially ruled out GUT scale strings. Such experiments include null results from lensing surveys \cite{Christiansen:2010zi}, gravitational wave background \cite{Abadie:2011fx} and burst searches \cite{Aasi:2013vna}, pulsar timing observations \cite{vanHaasteren:2011ni}, and observations of the cosmic microwave background \cite{Dvorkin:2011aj,Ade:2013xla} (see \cite{Copeland:2011dx} for a general review). Roughly speaking, these upper bounds imply $G \mu \lesssim 3 \times 10^{-8} - 3 \times 10^{-7}$, although all such bounds are model-dependent and subject to a variety of observational and astrophysical uncertainties, with more stringent bounds typically invoking additional assumptions. Future gravitational wave observatories (including Advanced LIGO [Laser Interferometer Gravitational Wave Observatory], LISA [Laser Interferometer Space Antenna] and Nanograv [North American Nanohertz Observatory for Gravitational Waves]) may achieve limits as low as $G \mu \sim 10^{-12}$ \cite{Olmez:2010bi}.

The situation for strings in modern string theory is somewhat different than it is for
GUT strings. In well-studied models of string theory, the compactification of extra dimensions involves manifolds possessing warped throat-like structures which redshift all characteristic energy scales compared to those in the bulk space. String theory contains multiple {\it effectively} one-dimensional objects collectively referred to here as superstrings. Any superstring we observe will have tension $\mu$ exponentially diminished from that of the Planck scale by virtue of its location at the bottom of the throat.  There is no known lower limit for $\mu$. The range of interest for microlensing is $10^{-14} < G \mu < 3 \times 10^{-7}$, with the lower limit set by the difficulty of observing optical microlensing of stars (finite source size versus lensing angle; finite duration versus lensing timescale) and the upper limit by the collective empirical tension bounds.

The lowered tension of superstring loops qualitatively alters their astrophysical fate compared to GUT strings. With much smaller $\mu$, superstring loops of a given size live longer. This has two important implications: the loops that dominate the lensing probability are small loops (sub-galactic rather than horizon-crossing), and were born before the matter era --- they are old. Because of their age, cosmic expansion has had sufficient time to dampen the initial peculiar motion of the loops, which allows them to cluster as matter perturbations grow \cite{Chernoff2007, Chernoff2009}. As such, below a critical tension $G \mu \sim 10^{-9}$, all superstring loops accrete along with cold dark matter.

The bottom line is that low-tension loops within the galaxy are over-abundant with respect to loops within the universe as a whole by roughly the same factor as cold dark matter is over-abundant within the Galaxy. By investigating a detailed model of the tension-dependent distribution of loops within our galaxy \cite{Chernoff2013}, it was found that the local enhancement of string loops has implications for microlensing searches, gravitational wave detection experiments and for pulsar timing measurements. In this paper, we concentrate on microlensing.

Previous work on cosmic string lensing has mainly assumed high string tensions, wherein multiple images of the same object may be resolved \cite{deLaix:1997dj, deLaix:1997jt, Uzan:2000xv, Shlaer:2005ry}. Numerical computations show that the lensing patterns of string loops can become quite involved \cite{Vachaspati1996}, with complicated caustics defining where multiple images may exist.

Based on the low string tensions required by recent cosmological observations, it is likely that only a microlensing signature can be detected optically, as low tensions imply small angular deflections \cite{Chernoff2007}. In a microlensing event, multiple unresolved images lead to an apparent doubling of flux over the duration of the event. The defining features of such an event are an achromatic doubling of flux that repeats many times with a characteristic
time-scale set by the loop period. Cosmic string microlensing is thus distinct from lensing in that it is a time domain measurement rather than an image signature, and is distinct from other astrophysical microlensing events because of its unique flux signature.

Stars are the ideal target for microlensing within the galaxy. They are numerous, and ever more capable time-domain surveys are being planned and carried out. Furthermore, the characteristic angular scale for microlensing is $8 \pi G \mu$, which is well-matched to the stellar size. For $G \mu=10^{-13}$ this angle is comparable to the angular size of a solar mass star at $10$ kpc. For $G \mu > 10^{-13}$ the stars are point-like targets. Microlensing of stars may be spatially correlated on the sky in a manner similar to that of normal lensing \cite{Huterer:2003ze}.

Current descriptions of rate calculations for microlensing events \cite{Kuijken:2007ma} neglect the galactic clustering effect, and updated results are in progress \cite{Chernoff2014}. It is likely that constraints on the parameter space of cosmic string loops can be determined from investigations of microlensing in current observational data.

It is important to derive a full understanding of the microlensing signature to be able to hunt for and identify the rare but meaningful events that may occur during a survey. In this paper, we begin a detailed exploration of the nature of microlensing for point-like sources, presenting the first numerical realizations of cosmic string microlensing. The framework of this dynamical calculation combines two elements that have not hitherto been melded but are equally important: the space curvature of the mass component of the bent loop and the deficit angle of the static string source. This opens the possibility of considering additional elements of interest, including substructure and discontinuities on the loop, and paves the way to making microlensing a viable search technique for current and upcoming large scale surveys such as PanStarrs, LSST, Gaia and WFIRST, and deep bulge surveys such as OGLE and MOA.

This paper is organised as follows. We begin with an overview of string lensing from infinite straight strings in Section \ref{sec:basics}. In Section \ref{sec:details}, we present the mathematical description of cosmic string loops and detail the formalism we use to propagate geodesics in the string spacetime. In Section \ref{sec:computation}, we describe the computational implementation of this formalism. We briefly describe the string configuration we use in Section \ref{sec:stringconfig}, before presenting the results of our numerical investigation in Section \ref{sec:results} and concluding in Section \ref{sec:conclusion}.

\section{String Lensing Basics}\label{sec:basics}

Lensing is a fundamental feature of general relativity. Curved spacetime can create multiple images with
varying magnification, shear and distortion. In {\it microlensing} the individual images of the source are
unresolved but the total brightness varies in time as the geometry of observer, lens and source evolve.  Refsdal \cite{Refsdal1964} calculated microlensing for a gravitating Newtonian point mass and concluded that significant brightness amplification was possible. Paczy\'{n}ski \cite{Paczynski1991} proposed utilizing microlensing to search for dark, massive objects contributing to the total mass density of the Galactic halo.

Vilenkin \cite{Vilenkin1984} was the first to discuss lensing by a cosmic string. Unlike a Newtonian point mass which curves spacetime, a straight string's positive energy density and negative pressure (along its length) conspire to leave spacetime flat. Its presence induces a deficit angle with size $\Delta = 8 \pi G \mu$ and creates a conical geometry.  There is no magnification, shear or distortion of a particular image, but there are multiple distinct paths for photons to travel from source to observer when the source, string and observer are all nearly aligned.  For GUT strings with $G\mu = 10^{-6}$ the typical angular splitting of images is $\sim 5$ arc seconds. Vilenkin noted that exact double images of cosmic objects like galaxies located behind the string might reveal the string's presence. Since that suggestion, the observational bounds (from CMB, gravity wave searches and pulsar timing) have tightened, constraining $G\mu \lesssim 3 \times 10^{-8} - 3 \times 10^{-7}$, and the expected splitting cannot be resolved. In addition, advances in string theory naturally yield superstrings, string-like entities which can have much smaller $\mu$. In this context, Chernoff and Tye \cite{Chernoff2007} suggested that one look for the transient change of the unresolved flux when a star passes behind the string. A search for superstrings by string microlensing is conceptually similar to the search for Newtonian masses by way of normal microlensing.

\begin{figure}
  \centering
  \includegraphics[width=\columnwidth]{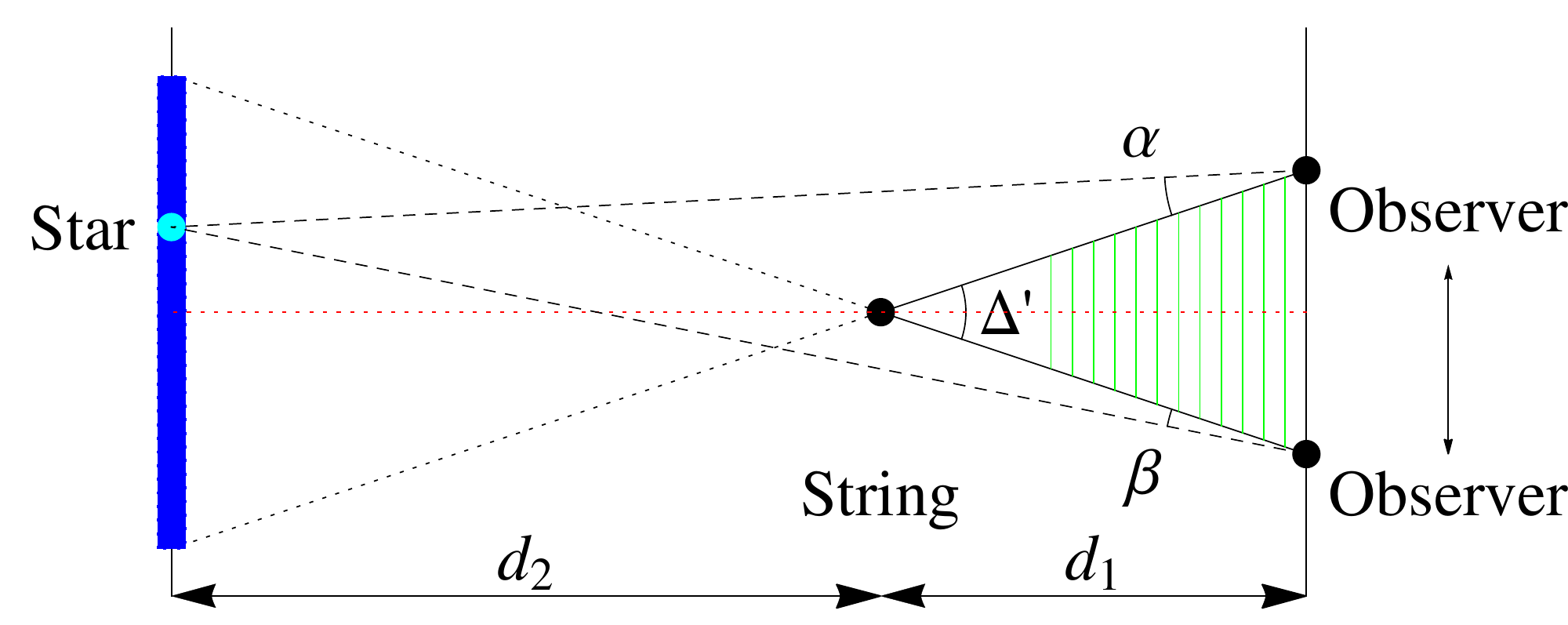}
  \caption{The black dot at the center is a straight string with tension $\mu$ and intrinsic deficit angle $\Delta = 8 \pi G \mu$. An angular section of Minkowski space has been aligned with the observer and removed (green hatched region).
This makes clear that certain points can transmit light to the observer by both clockwise and counterclockwise circumnavigation of the string. The red dotted line is the symmetrical line from observer to string.  The string pierces the plane formed by the star on the left (blue dot) and the observer on the right (black dots) at angle $\theta$ (not drawn). The wedge has projected deficit angle $\Delta' = \Delta \sin \theta$.  When the star at distance $d_2$ lies along the blue strip subtended by the dotted lines there exist two lines of sight for the observer to see the star. The total flux is very nearly double the flux of a single image and the angular separation of the pair of images is $\delta \phi = \alpha + \beta$.  }
  \label{fig:lensing}
\end{figure}

The lensing geometry for static, straight infinite strings is illustrated in Figure \ref{fig:lensing}. The observer-lens distance is $d_1$, the star-lens distance is $d_2$ and the observer line of sight to the star is ${\hat n}$. Lensing requires close alignment of star, string and observer to order $\Delta$. We will work to lowest non-vanishing order in $\Delta$.

The string unit vector ${\hat s}$ forms an angle $\theta$ with respect to the line of sight ($\cos \theta = {\hat n} \cdot {\hat s}$).  The wedge spans the projected deficit angle $\Delta' = \Delta \sin \theta$,
is removed from the plane (green hatched region) and opposite sides identified.
When the star lies anywhere along the blue strip on the left, two lines of sight exist for the observer to see the star. Let the displacement of the star from line of symmetry in the plane be $h$, and write ${\tilde d}_1=d_1/(d_1+d_2)$, ${\tilde d}_2=d_2/(d_1+d_2)$, ${\tilde h}=2 h/(d_2 \Delta')$. The range of microlensing is $|{\tilde h}|<1$.

The angle of separation between the images is
\begin{align}
 \delta \phi = \alpha + \beta = {\tilde d}_2 \Delta' + O(\Delta^3),
\end{align}
the path length difference is
\begin{align}
\frac{\delta d}{d_1+d_2} =
\frac{1}{2} {\tilde d}_1 {\tilde d}_2 {\tilde h} (\Delta')^2 + O(\Delta^4)
\end{align}
and the fractional change in energy for photons in this geometry is
\begin{align}
\frac{\delta \nu}{\nu}
= \Delta \gamma {\hat n} \cdot \left( {\vec v} \times {\hat s} \right)
\end{align}
where $\gamma$ is the relativistic factor for the string \cite{Kaiser:1984iv, Vachaspati1986, Vilenkin1986, Stebbins1987}. During a microlensing event, geodesics must have a distance of closest approach to the string $r$ in the range
\begin{align}
  |r| \lesssim \frac{d_1 d_2}{d_1 + d_2} \Delta^\prime \,.
\end{align}
The microlensing event will have a duration of
\begin{align}
  \delta t \approx \frac{2 d_1 d_2}{d_1 + d_2} \frac{\Delta^\prime}{|\vec{v} \cdot (\hat{n} \times \hat{s})|} \gtrsim \frac{2 d_1 d_2}{d_1 + d_2} \Delta^\prime \label{eq:timing}
\end{align}
as the string sweeps through the line of sight.

The most numerous string loops today are expected to have a length $\ell \sim \Gamma G \mu \tau$, where $\tau$ is the age of the universe, and $\Gamma \sim 50-100$. Such a loop has a typical curvature scale $\sim 1/\ell$ and a bounding box of side length $\sim \ell/4$. Microlensing caused by the deficit angle is possible as long as the distance of closest approach satisfies $r \ll \ell$; otherwise the loop's mass will lens (or microlens) in a traditional Newtonian fashion. The first possibility requires that the source (observer) distance must be less than $\ell/\Delta^\prime$, which is comparable to the Hubble scale today. The infinite straight string results should provide a valid approximation as long as the characteristic string scale is $\sim \ell$. Of course, an individual segment of string may have a much smaller radius of curvature, particularly near a cusp, where these results become more approximate. One of the purposes of this paper is to compare these analytical results to numerical simulations for microlensing in cosmic string loops.

\section{Gravitational Wave Emission and Geodesic Propagation}\label{sec:details}

In this section, we present the formalism used to calculate the metric in the string spacetime. We also describe the geodesic equation formulation that we employ.

\subsection{String Dynamics}
We begin with a brief overview of string dynamics following Vilenkin and Shellard \cite{Vilenkin1994}. The string is defined by its location $x^\mu(\xi^a)$, where $\xi^0 = \tau$ and $\xi^1 = \sigma$ are the worldsheet coordinates. We work with the Nambu-Goto string action
\begin{align}
  S = - \mu \int d^2 \xi \sqrt{-\gamma}
\end{align}
where $\mu$ is the string tension, and $\gamma_{ab} = g_{\mu \nu} \partial_a x^\mu \partial_b x^\nu$ is the induced metric on the worldsheet.

Working in flat space, the equation of motion for the string is given by
\begin{align}
  \ddot{x}^\mu - x^{\mu \prime \prime} = 0
\end{align}
where $\dot{x}^\mu = \partial x^\mu / \partial \tau$ and $x^{\mu \prime} = \partial x^\mu / \partial \sigma$, and we use the conformal gauge
\begin{align}
  \dot{x} \cdot x^\prime = 0, \qquad \dot{x}^2 + x^{\prime 2} = 0.
\end{align}
We further simplify the gauge by choosing $\tau = t$, the time coordinate. This allows us to write the string trajectory as a three-vector $\vec{x} (\sigma, t)$, subject to the conditions
\begin{align}
  \dot{\vec{x}} \cdot \vec{x}^\prime = 0, \qquad \dot{\vec{x}}^2 + \vec{x}^{\prime 2} = 1, \qquad \ddot{\vec{x}} - \vec{x}^{\prime \prime} = 0.
\end{align}
The general solution to these equations is
\begin{align}
  \vec{x}(\sigma, t) = \frac{1}{2} \left[ \vec{a} (\sigma - t) + \vec{b} (\sigma + t) \right]
\end{align}
subject to the conditions
\begin{align}
  \vec{a}^{\prime 2} = \vec{b}^{\prime 2} = 1
\end{align}
which imply that the tangent vectors live on a unit sphere. The energy of the string is simply $E = \mu L$ for invariant string length $L$, and the stress-energy tensor for the string is given by
\begin{align}
  T^{\mu \nu} (\vec{r}, t) = \mu \int d \sigma (\dot{x}^\mu \dot{x}^\nu - x^{\mu \prime} x^{\nu \prime}) \delta^{(3)} (\vec{r} - \vec{x}(\sigma, t)).
\end{align}

\subsection{Metric Perturbations}
We work in linearized gravity, with $g_{\mu \nu} = \eta_{\mu \nu} + h_{\mu \nu}$ and metric signature $(-,+,+,+)$. The linearized Einstein equation is given by
\begin{align}
  \square h_{\mu \nu} = - 16 \pi G S_{\mu \nu}
\end{align}
where
\begin{align}
  S_{\mu \nu} = T_{\mu \nu} - \frac{1}{2} \eta_{\mu \nu} \tensor{T}{^\sigma_\sigma}
\end{align}
and we work in harmonic gauge
\begin{align}
  \partial_\nu \tensor{h}{^\nu_\mu} = \frac{1}{2} \partial_\mu \tensor{h}{^\sigma_\sigma}.
\end{align}

To invert the linearized Einstein equation, we use the Green's function method following Damour and Buonanno \cite{Buonanno:1998is}. The inversion yields
\begin{align}
  h_{\mu \nu} (\vec{r}, t) &= \int d^2 \sigma \; 8 G \mu \; F_{\mu \nu} \theta(t - x^0) \delta((r-x)^2)
\end{align}
for the metric perturbation, where
\begin{align}
  F_{\mu\nu} = \dot{x}_\mu \dot{x}_\nu - x_\mu^\prime x_\nu^\prime + \eta_{\mu \nu} x^{\sigma \prime} x_\sigma^\prime.
\end{align}
Here, $r$ and $x$ in the delta function refer to the four-vectors, where $r^\mu = (t, \vec{r})$. Defining $\Omega^\mu = r^\mu - x^\mu (\sigma, \tau)$, the retarded time needed in integrating out the delta function is the retarded solution to
\begin{align}
  \eta_{\mu \nu} \Omega^\mu \Omega^\nu = 0
\end{align}
which yields
\begin{align}
  \tau = t - |\vec{r} - \vec{x}(\sigma, \tau)|.
\end{align}
We call the solution to this equation $\tau_{ret}$, the retarded time. Integrating over the delta function, the metric perturbation becomes
\begin{align}
  h_{\mu \nu} (\vec{r}, t) &= \int d \sigma \; 4 G \mu \; \frac{F_{\mu \nu} (\sigma, \tau_{ret})}{|\Omega^\mu \dot{x}_\mu (\sigma, \tau_{ret})|}. \label{eq:metpert}
\end{align}
Note that $\Omega^\mu \dot{x}_\mu$ is negative when evaluated on the retarded time, except on the string or on a line of cusp radiation, where it vanishes and the integral diverges logarithmically. If desired, one can avoid this situation by implementing an over-retarded time with $\tau = t - \sqrt{|\vec{r} - \vec{x}(\sigma, \tau)|^2 + \epsilon^2}$, which regulates the denominator.
\begin{align}
\Omega^\mu \dot{x}_{\mu} = (\vec{r} - \vec{x}) \cdot \dot{\vec{x}} - \sqrt{|\vec{r} - \vec{x}|^2 + \epsilon^2} < 0
\end{align}
A reasonable size for $\epsilon$ is the distance of closest approach at which the metric perturbation becomes nonlinear. In the numerical part of this work, $\epsilon$ was set to zero for all practical purposes.

Calculating the average field a long way away from the loop, one finds
\begin{align}
  \left< h_{\mu \nu} (\vec{r}) \right> &= \frac{2}{L} \int_0^{L/2} h_{\mu \nu} (\vec{r}, t) dt
\nonumber\\ &= \frac{2 GM}{|\vec{r}|} \textrm{diag} (1,1,1,1) \label{eq:averagefield}
\end{align}
which corresponds to the linearized Schwarzschild metric for a loop of mass $M = E = \mu L$.

\subsection{Metric Derivatives}
The reason for using Damour and Buonanno's approach is that it allows for metric derivatives to be straightforwardly calculated. A single partial derivative acting on the metric perturbation is
\begin{align}
  \partial_\alpha h_{\mu \nu} (r) &= \int d\sigma d\tau \; 8 G \mu \; F_{\mu \nu}(\sigma, \tau) \theta(r^0 - \tau) \partial_\alpha \delta(F)
\end{align}
where we let $F = (r-x(\sigma, \tau))^2 = \eta_{\mu \nu} \Omega^\mu \Omega^\nu$. Note that the derivative acting on the step function yields a product of delta functions at two separate locations, which vanishes (except on the string). The derivative acting on the delta function yields
\begin{align}
  \partial_\alpha \delta(F) = \partial_\alpha F \delta^\prime (F) = \frac{\partial_\alpha F}{\partial F/\partial \tau} \frac{\partial \delta (F)}{\partial \tau}
\end{align}
by repeated use of the chain rule. Integrating by parts and noting that derivatives of $F$ can be simply expressed in terms of $\Omega^\mu$ as
\begin{align}
  \frac{\partial F}{\partial \tau} = -2 \Omega^\mu \dot{x}_\mu\,, \qquad
  \partial_\alpha F = 2 \Omega_\alpha\,,
\end{align}
leads to Damour and Buonanno's Eq. (2.40),
\begin{align}
  \partial_\alpha h_{\mu \nu} (r) &= \int d\sigma \frac{4 G \mu}{|\Omega^\mu \dot{x}_\mu (\sigma, \tau)|} \frac{\partial}{\partial \tau} \left[F_{\mu \nu}(\sigma, \tau) \frac{\Omega_\alpha}{\Omega^\mu \dot{x}_\mu} \right] \label{eq:metpertprime}
\end{align}
where all expressions should be evaluated at $\tau_{ret}$. The $\tau$ derivative yields
\begin{align}
  \frac{\partial}{\partial \tau} \left[F_{\mu \nu} \frac{\Omega_\alpha}{\Omega^\mu \dot{x}_\mu} \right] &=
  \dot{F}_{\mu \nu} \frac{\Omega_\alpha}{\Omega^\mu \dot{x}_\mu}
  -
  F_{\mu \nu} \frac{\dot{x}_\alpha}{\Omega^\mu \dot{x}_\mu}
\nonumber\\& \qquad
  -
  F_{\mu \nu} \frac{\Omega_\alpha}{(\Omega^\mu \dot{x}_\mu)^2} \Lambda
\end{align}
where we define $\Lambda = \partial_\tau (\Omega^\mu \dot{x}_\mu) = \Omega^\mu \ddot{x}_\mu - \dot{x}^\mu \dot{x}_\mu$.

Extending this approach, we can compute the second derivative of the metric, which will be needed to compute curvature invariants. Starting from
\begin{align}
  \partial_\alpha \partial_\beta h_{\mu \nu} (r) &= \partial_\beta \int d\sigma d\tau \; 8 G \mu \; F_{\mu \nu} \theta(r^0 - \tau) \partial_\alpha \delta(F) \,,
\end{align}
we follow the same idea as for the first derivative: evaluate all the derivatives before integrating out the delta function. The resulting expression is as follows.
\begin{widetext}
\begin{align}
  \partial_\alpha \partial_\beta h_{\mu \nu} (\vec{r}, t)
  &= \int d\sigma \frac{4 G \mu}{|\Omega^\mu \dot{x}_\mu|} \frac{\partial}{\partial \tau} \left[F_{\mu \nu} \frac{\eta_{\alpha \beta}}{\Omega^\mu \dot{x}_\mu}
  - F_{\mu \nu} \frac{\dot{x}_{\beta} \Omega_{\alpha} + \dot{x}_{\alpha} \Omega_{\beta}}{(\Omega^\mu \dot{x}_\mu)^2}
  + \dot{F}_{\mu \nu} \frac{\Omega_\alpha \Omega_\beta}{(\Omega^\mu \dot{x}_\mu)^2}
  - F_{\mu \nu} \frac{\Omega_\alpha \Omega_\beta}{(\Omega^\mu \dot{x}_\mu)^3} \Lambda
  \right]
\end{align}
All expressions here should be evaluated at $\tau_{ret}$. Evaluating the derivatives, the final result is
\begin{align}
  \partial_\alpha \partial_\beta h_{\mu \nu} (\vec{r}, t)
  &= \int d\sigma \; \frac{4 G \mu \; \Sigma_{\mu \nu \alpha \beta}}{|\Omega^\mu \dot{x}_\mu (\sigma, \tau_{ret})|^3} \label{eq:metpertprimeprime}
\end{align}
where
\begin{align}
\Sigma_{\mu \nu \alpha \beta} &= F_{\mu \nu} \left[
  - 2 \ddot{x}_{(\alpha} \Omega_{\beta)}
  + 2 \dot{x}_{\alpha} \dot{x}_{\beta}
  - \eta_{\alpha \beta} \Lambda
  + 6 \frac{\dot{x}_{(\alpha} \Omega_{\beta)}}{\Omega^\mu \dot{x}_\mu} \Lambda
  + 3 \frac{\Omega_\alpha \Omega_\beta}{(\Omega^\mu \dot{x}_\mu)^2} \Lambda^2
  - \frac{\Omega_\alpha \Omega_\beta}{\Omega^\mu \dot{x}_\mu} \dot{\Lambda}
  \right]
\nonumber \\ &{} \qquad
  + \dot{F}_{\mu \nu} \left[
  \eta_{\alpha \beta} \Omega^\lambda \dot{x}_\lambda
  - 2 (\dot{x}_{\beta} \Omega_{\alpha} + \dot{x}_{\alpha} \Omega_{\beta})
  - 3 \frac{\Omega_\alpha \Omega_\beta}{\Omega^\mu \dot{x}_\mu} \Lambda
  \right]
  + \ddot{F}_{\mu \nu} \Omega_\alpha \Omega_\beta
\end{align}
and $\dot{\Lambda} = \partial_\tau \Lambda = \Omega^\mu \dddot{x}_\mu - 3 \ddot{x}^\mu \dot{x}_\mu$.
\end{widetext}

\subsection{Geodesic Equation}
We are interested in ray-tracing null geodesics through the perturbed spacetime. The geodesic equation is given by
\begin{align}
  \frac{\partial^2 x^\mu}{\partial \lambda^2} + \tensor{\Gamma}{^\mu_{\sigma \lambda}} \frac{\partial x^\sigma}{\partial \lambda} \frac{\partial x^\lambda}{\partial \lambda} = 0
\end{align}
where $x^\mu$ describes the position of a photon in spacetime, and $\lambda$ is an affine parameter. This equation can be integrated directly, but there exist better approaches. Because of the mass shell constraint $p^2 = 0$ (where $p^\mu = \partial x^\mu / \partial \lambda$, with affine parameter $\lambda$ chosen such that $p^\mu$ is the four-momentum of the photon), only three components of the geodesic equation actually need to be integrated. Furthermore, the affine parameter can be disposed of by using coordinate time, as no horizons are present in the spacetimes under consideration. An efficient method of integrating the geodesic equation that takes advantage of these properties is described by Hughes \textit{et al.} \cite{Hughes:1994ea}, which does not require time derivatives of the metric components and thus saves computational time.

Consider the metric written in the ADM decomposition.
\begin{align}
  ds^2 = - \alpha^2 dt^2 + \gamma_{ij} (dx^i + \beta^i dt) (dx^j + \beta^j dt)
\end{align}
With the metric written in this manner, we can write the geodesic equation as follows.
\bse \label{eq:geodesics}
\begin{align}
  \frac{dx^i}{dt} &= \gamma^{ij} \frac{p_j}{p^0} - \beta^i \label{eq:geodx}
\\
  \frac{dp_i}{dt} &= - \alpha \alpha_i p^0 + \tensor{\beta}{^k_{,i}} p_k - \frac{1}{2} \tensor{\gamma}{^{jk}_{,i}} \frac{p_j p_k}{p^0}
\\
  p^0 &= \frac{1}{\alpha} \sqrt{\gamma^{ij} p_i p_j}
\end{align}
\ese
The variables being integrated are $p_i$ and $x^i$; the energy $p^0$ is algebraically determined at each step, ensuring the mass-shell condition. The inverse spatial metric $\gamma^{ij}$ is used to raise and lower indices on $\beta^i$. Note that the affine parameter has vanished, and that no time derivatives of the metric are needed. This formulation is completely general, but fails near horizons, where the time coordinate becomes problematic. This system of equations can be specialized to the linear approximation if desired. However, the amount of extra computational time required to numerically integrate the full equations compared to the linearized equations is negligible.

%We now specialize to the linearized approximation. In terms of the metric perturbations, we have the following to linearized order in $h$.
%\begin{align}
%  \gamma_{ij} &= \delta_{ij} + h_{ij}
%\\
%  \gamma^{ij} &= \delta^{ij} - h^{ij} + O(h^2)
%\\
%  \beta^i &= \delta^{ij} \beta_j + O(h^2) = \delta^{ij} h_{jt}
%\\
%  \alpha^2 &= \beta^i \beta^j \delta_{ij} + 1 - h_{tt} = 1 - h_{tt} + O(h^2)
%\end{align}
%Here, $h^{ij}$ has its indices raised with the flat Euclidean spatial metric. The metric derivatives then follow straightforwardly.
%\begin{align}
%  \alpha \alpha_{,i} &= - \frac{1}{2} h_{tt,i}
%\\
%  \tensor{\beta}{^k_i} &= \delta^{kj} h_{jt,i}
%\\
%  \tensor{\gamma}{^{jk}_{,i}} &= - \delta^{jm} \delta^{kn} h_{mn,i}
%\end{align}

\section{Computational Tools}\label{sec:computation}

In this section, we describe the computational approach that we employ to calculate metric perturbations and their derivatives, to integrate geodesics, and to solve the geodesic equation as a boundary value problem. We also include a detailed analysis of the approximations included in our calculations.

The formalism that we use to calculate geodesics in the microlensing context is similar to that developed by de Laix and Vachaspati \cite{Vachaspati1996} to investigate the lensing properties of cosmic strings. However, there are a number of differences. In particular, we work with finite source and observer distance, rather than placing them at infinite distances. Furthermore, we do not work in a thin lens approximation. Finally, because we trace the full geodesic rather than just looking at the deflection angle, we can investigate features along the geodesic that allow us to understand phenomena in the observables, and also compare theoretical predictions to details of the geodesics such as distance of closest approach.

To calculate the metric perturbation and its derivatives at a given point in spacetime, we make use of Eqs. \eqref{eq:metpert}, \eqref{eq:metpertprime} and \eqref{eq:metpertprimeprime}. These formulas all include an integral over the string loop. The required components are stored as a vector and integrated together.

At each point on the string loop, the retarded time needs to be evaluated. As the implicit equation describing the retarded time is strictly monotonic (except for exactly on a cusp where it is stationary), this is reasonably straightforward. For points on the string close to a cusp, the derivative of the implicit equation becomes very small near the root, and so a combination of Newton's method and a bisection method are employed. We use a tolerance $\delta \tau$ for the accuracy of the retarded time calculation.

Once the retarded time has been evaluated, it is straightforward to evaluate the integrands for the metric perturbations and their derivatives. The integral is then summed using a psuedospectral method. As the string loops we considered were smooth and periodic, it is highly efficient to evaluate the integral by a bisection method, using a rectangular approximation. We begin by evaluating the integral with 32 divisions, and compare it to 64 divisions in order to estimate relative and absolute accuracy. The number of bisections is doubled until the relative and absolute tolerances are satisfied. Due to the limits of long integers, we demand that at most 30 bisections be allowed, although such large numbers are only typically needed when evaluating points very close to the string. While integrating over the loop, the shortest distance to the loop (in the form $t - \tau_{ret}$) is recorded. Using this method of integration precludes the investigation of strings with kinks, which are not appropriately continuous. In principle, the method also breaks down if a cusp is encountered (due to the discontinuity), but as cusps only form for an instant in time, the string can always be taken to be smooth.

In order to understand how an object is microlensed, we want to know the null geodesic(s) that start at the source and end at the observer at a given time. This specifies a geodesic as a boundary value problem. For the string spacetimes we are considering, there will always be at least one solution (a deviation from the Euclidean geodesic), and the possibility of microlensing suggests that there will sometimes be multiple solutions.

We solve the geodesic equation as an initial value problem by selecting an observer position (three boundary conditions) and photon arrival direction (two angles describing a unit vector). The third component of the momentum sets the arrival energy, which we normalise to one. Coordinate time is used as the integration parameter, and so an initial time must also be selected. The initial value problem is thus specified in terms of 5 boundary conditions and one parameter. To integrate the geodesics, we make use of the GNU Scientific Library (GSL) \cite{GSL} ODE integration routines. We integrate the position and $p_i$ components backwards in time for a predetermined period of time, using the Runge-Kutta-Fehlberg (4, 5) method. The accuracy on each integration step is specified in terms of relative and absolute error tolerances, and an adaptive step size is used.

When searching for a geodesic that connects the source and observer, there are three parameters that can be varied: two initial conditions which control the angle at which the photon arrives at the observer, and one parameter which describes the time the source emitted the photon (and thus how far back to integrate the geodesic). As we are working with linearized gravity about a Minkowski background, the flat space geodesic that connects the source and observer provides a good initial guess for these parameters. In order to improve a guess, we consider geodesics with angles that slightly vary in orthogonal directions and also propagate the geodesic a little further backward in time, constructing three basis vectors. We then calculate the Euclidean deviation from the beginning of the geodesic to the source, and decompose that deviation in terms of the three basis vectors. The corresponding modifications are then made to the angles and timing, and the process repeated until the geodesic lands within some tolerance of the source. This method is computationally intensive: each iteration requires three geodesics to be computed. Nonetheless, the approach typically led to exponential convergence up to the point where the tolerances were insufficiently tight to allow further convergence.

Iterating this procedure for arrival times across one period of the string loop obtains one geodesic per arrival time. For our initial searches, we broke the string period up into 1000 timesteps. In order to obtain multiple geodesics at a given arrival time, we searched for geodesics that arrived one time step apart and whose angle at the observer jumped significantly. Such pairs were always found to have local minima in the distance of closest approach. We then used the angles for these geodesics modified by the appropriate angular velocity as initial guesses for time steps moving forward or backward in time appropriately in order to obtain multiple geodesics arriving at a given time. Due to rapid changes occurring as photons approach a string segment very closely, we used increasingly small timesteps to obtain new geodesics. This method of identifying microlensing solutions relies on each family of smoothly-connected geodesics being the only solution at some point in the evolution. For microlensing events where a microlensed image appears and then disappears while the primary image is always visible, this method is unlikely to identify the second set of solutions. A source might be microlensed twice by two different lengths of string that lie close to the observer-source line of sight, an intrinsically low probability event. Generally speaking, events with more than two images will not be identified with this method.

\subsection{Approximations}
Our approach to calculating metric perturbations is fundamentally limited by the linearized gravity approximation. In our numerical implementation, a number of approximations are made, mainly involving accuracy tolerances. Our approach to our numerical approximations is to ensure that the numerical precision is superior to the linear approximation. In this section, we estimate the magnitude of each error, and calculate appropriate tolerances.

The linearized gravity approximation drops terms of order $h^2$, where $h$ is the metric perturbation. From Eq. \eqref{eq:metpert}, the metric perturbation in the linear regime is
\begin{align}
  h_{\mu \nu} = 4 G \mu \int_0^L d\sigma \frac{F_{\mu \nu}}{|\Omega^\mu \dot{x}_\mu|} \,.
\end{align}
Heuristically, the error from the linear approximation is roughly $\delta h \sim h^2 \sim (G \mu)^2$, and so we aim for the numerical errors to also be of this order.

We would like to know when the linear perturbation becomes nonlinear. Very roughly, the numerator $F_{\mu \nu} \sim O(1)$, while the denominator $|\Omega^\mu \dot{x}_\mu| \sim |\vec{r} - \vec{x}|$. Let the distance of closest approach be $r_{min}$. The segment of string nearest this distance of closest approach will dominate the metric perturbation integral. As a crude estimate, model that segment of string as a straight segment of string of length $\alpha L$, with its center offset from the point of interest by the distance of closest approach. We have
\begin{align}
  h & \sim 4 G \mu \int_{-\alpha L/2}^{\alpha L/2} \frac{dh}{\sqrt{h^2+r_{min}^2}}
\nonumber \\
  & \sim 8 G \mu \ln \left( \frac{\alpha L}{r_{min}}\right) + O\left(\frac{r_{min}^2}{\alpha^2 L^2}\right) \,, \label{eq:happrox}
\end{align}
displaying the typical logarithmic divergence. For $h \sim 1$, we find
\begin{align}
  r_{min} \sim \alpha L e^{-1/8 G \mu} \,,
\end{align}
showing that the linear approximation should be valid up to distances very, very close to the string.

When calculating the metric perturbations, the integration routine estimates relative and absolute errors. Writing $h = h_{true} + \delta h_{integration}$, the relative error is $\delta h_{integration} / h$ and the absolute error is $\delta h_{integration}$. Comparing to the linear approximation, tolerances of $G\mu$ for the relative error and $(G \mu)^2$ for the absolute error are reasonable. As we desire the numerical errors to be subdominant compared to the linear approximation, we set the tolerances to be two orders of magnitude smaller again.

The metric perturbation is also sensitive to errors in the retarded time calculation. Roughly speaking, the error in the metric perturbation can be estimated as
\begin{align}
  \delta h \sim h \left(\frac{\delta \tau}{L} + \frac{\delta \tau}{r_{\star}}\right)
\end{align}
where $r_{\star}$ is the closest retarded distance to the string from that point in spacetime. The first term comes from errors in the numerator $F_{\mu \nu}$, while the second term comes from errors in the denominator $\Omega^\mu \dot{x}_\mu$. The appropriate conditions for the tolerance in the retarded time are then $\delta \tau < G \mu L$, $\delta \tau < G \mu r_{\star}$. In practice, because root finding using Newton's method is very cheap when a good estimate for the root is known (such as calculated from the previous segment of loop considered), we demanded $\delta \tau < (G \mu)^2 L / 100$, which satisfies both requirements up to $r_{\star}$ two orders of magnitude closer than we expect is required for microlensing to exist.

We now come to integrating the geodesic. At each integration step, two errors are calculated: the absolute and relative error. The total absolute error accumulated across the geodesic will be the number of steps times the absolute error per step. As we do not know a priori how many steps will be taken, we thus demand an absolute error of zero. In order to understand the relative error, we look at Eq. \eqref{eq:geodx} over a finite difference step. The error in computing $\Delta x^i$ from the linear approximation is $O(h^2)$, and so the tolerance on the relative error should be $G \mu$. Again, we set this to be a couple of orders of magnitude lower in order to make the numerical errors subdominant to the linear approximation.

The total error in the location at which the geodesic lands arising from the linear approximation is then $\sim G \mu (d_1 + d_2)$, where $d_1 + d_2$ is the Euclidean distance from the source to the observer. Again, we demand that the numerical precision is a couple of orders of magnitude better than this.

We see that essentially all tolerances scale with $G \mu$. As such, working at smaller $G \mu$ becomes more challenging, as satisfying the desired tolerances becomes more computationally demanding.

\section{Stringy Spacetimes}\label{sec:stringconfig}

In this section, we introduce the string loop configuration that we used, and investigate the spacetime that it generates.

\begin{figure}[tb!]
  \centering
  \includegraphics[width=0.8\columnwidth]{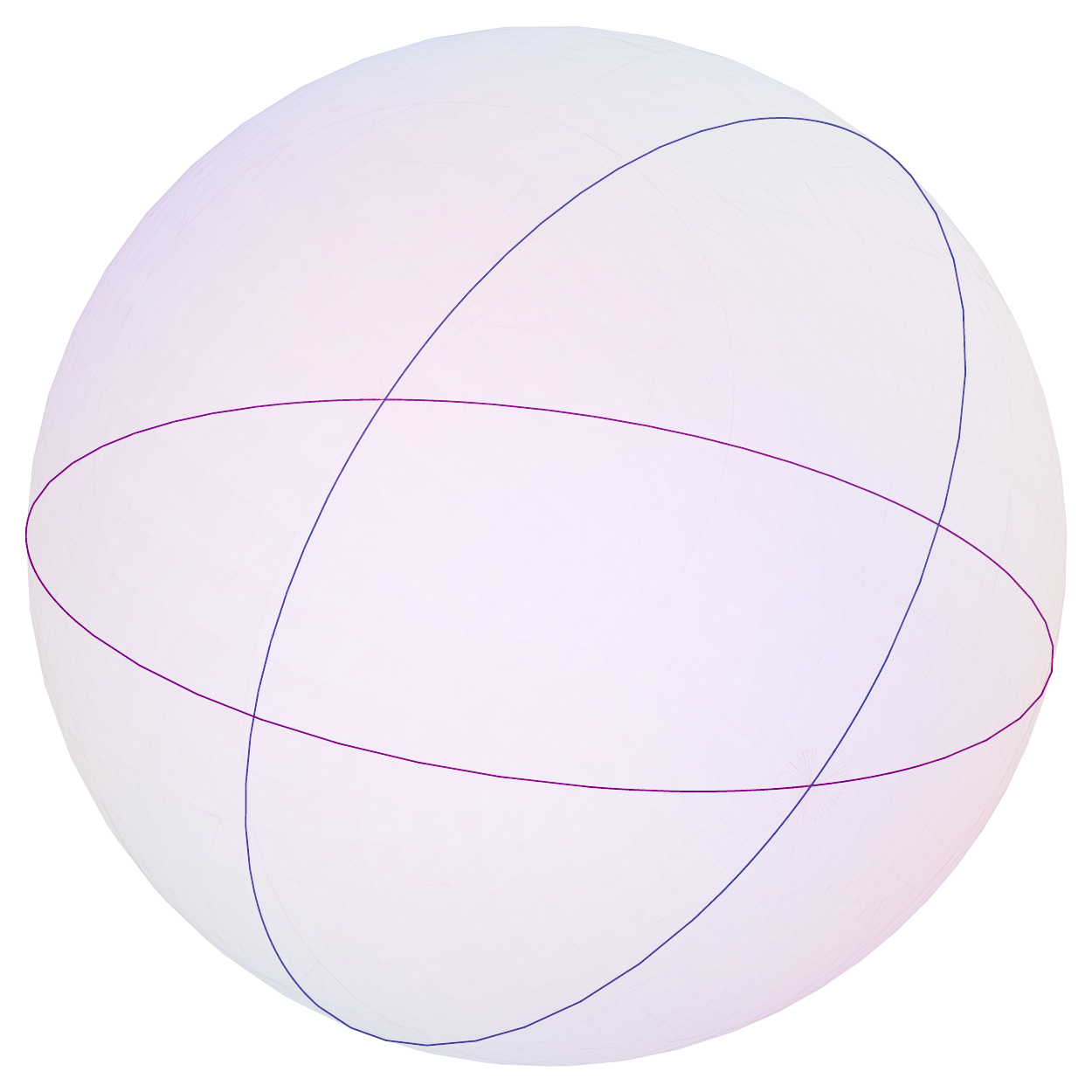}
  \caption{The tangent sphere representation of our string. One great circle is traversed twice for every time the other is traversed once.}
  \label{fig:tangentsphere}
\end{figure}

\begin{figure*}[p]
  \centering
  \includegraphics[width=0.6\textwidth]{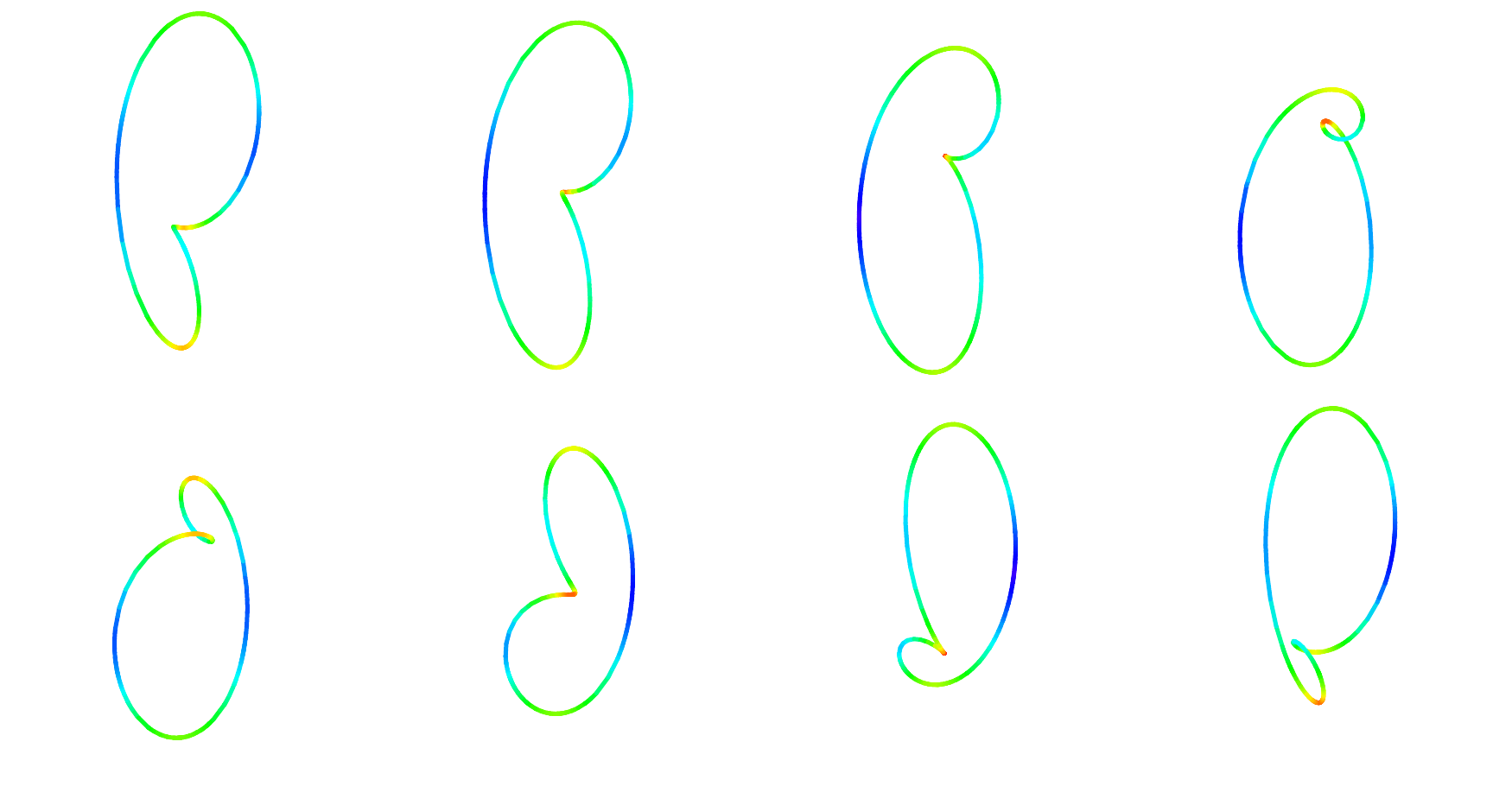}
  \caption{The loop considered in this paper in snapshots in time. The stills are equally spaced throughout the period, moving left to right, top to bottom. Note that these are plots of the position of the string, not the retarded position as seen by an observer from a string that ``shines''. The colour-coding displays the string's velocity at that point, with blue being slowest, and red approaching the speed of light. Images three and seven are at times near when a cusp forms.}
  \label{fig:loops}
\end{figure*}

For our goal of investigating string microlensing, we desired the simplest loop configuration we could find. More complicated loop configurations would be more computationally demanding when calculating the metric perturbations and their derivatives, and would complicate the interpretation of the results. As such, we chose to work with the Burden loop configuration \cite{Burden:1985md}, defined as follows.
\bse
\begin{align}
  \vec{a}(\xi) &= \frac{1}{\alpha} \big( \sin(\alpha \xi), \, 0, \, \cos (\alpha \xi)\big)
\\
  \vec{b}(\xi) &= \frac{1}{\beta} \big( \cos(\psi) \sin(\beta \xi), \, \sin(\psi) \sin(\beta \xi), \, \cos (\beta \xi)\big)
\end{align}
\ese
Here, $\alpha = 2 \pi m / L$ and $\beta = 2 \pi n / L$ for relatively prime integers $m$ and $n$, and $\psi$ is an arbitrary angle. When one of $m$ or $n$ is unity, this configuration has no self-intersections (for $0 < \psi < \pi$) or kinks, but possesses a cusp. The period of the loop is $T = L/2mn$. The tangent sphere representation is two great circles, with $n$ and $m$ revolutions each as $\sigma$ varies from 0 to $L$. For this work, we used $L = 2\pi$ in arbitrary units, $m = 1$, $n = 2$, and $\psi = \pi/3$. The tangent sphere for our loop is shown in Figure \ref{fig:tangentsphere}, and some images of the loop configuration through its period are presented in Figure \ref{fig:loops}. The string loop sits roughly along the $x$-$z$ plane.

\begin{figure*}[p]
  \centering
  \includegraphics[width=\textwidth]{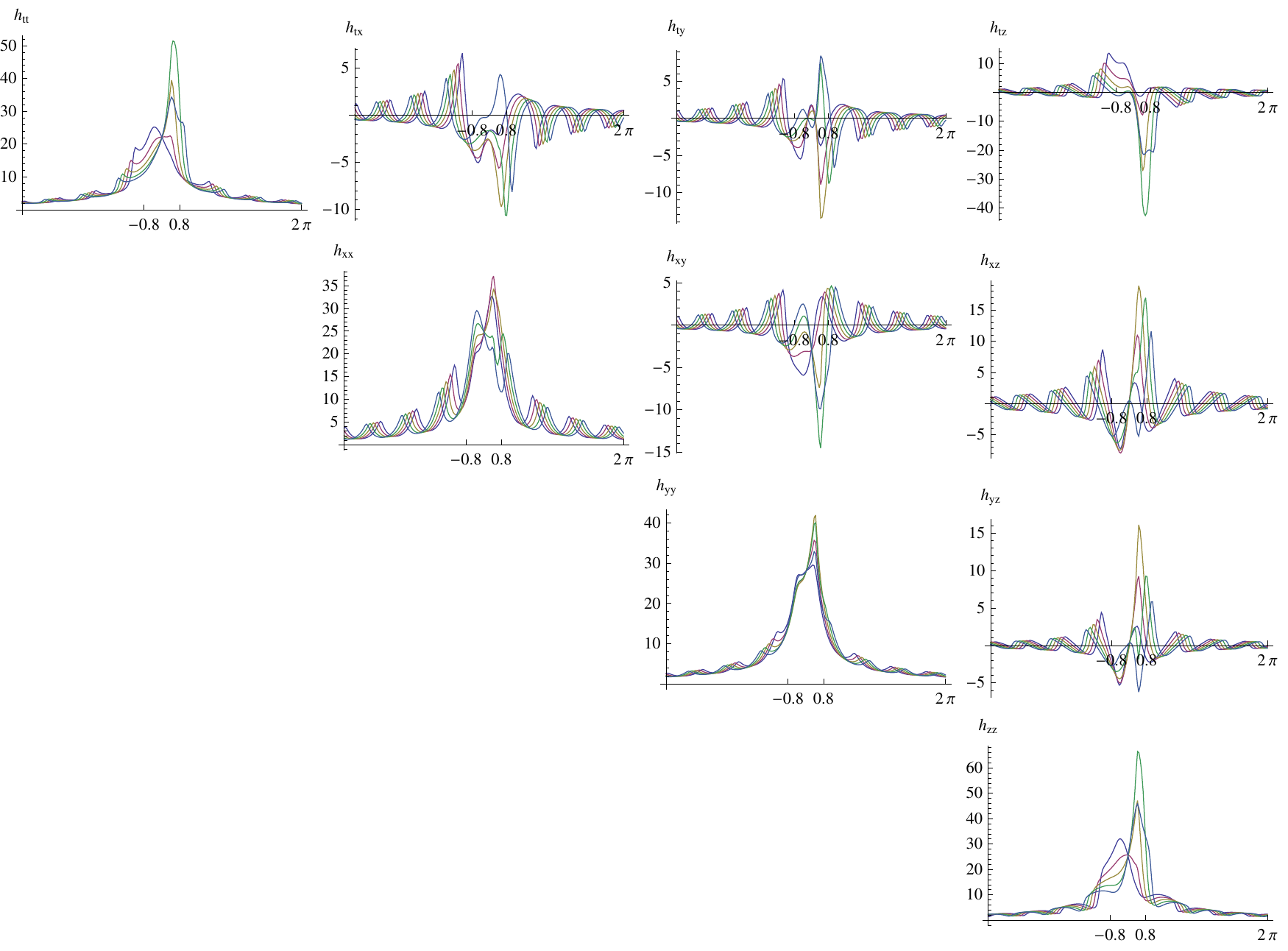}
  \caption{Plots of the metric perturbation components (divided by $G \mu$) along the $x$ axis, ranging from $-L$ to $L$ (where $L = 2 \pi$). Five times have been plotted, with 1/10th of the period passing between each curve. For reference, the string is contained in the bounds $-0.75 \le x \le 0.75$.}
  \label{fig:metdata}
\end{figure*}

Despite the simple nature of this particular loop configuration, the metric solution it sources is still highly complicated. To demonstrate this, we plot the metric components along the $x$ axis for different times in Figure \ref{fig:metdata} (there is nothing special about the $x$ axis; these plots are simply intended to be indicative of behavior).

\begin{figure}[tb!]
  \centering
  \includegraphics[width=\columnwidth]{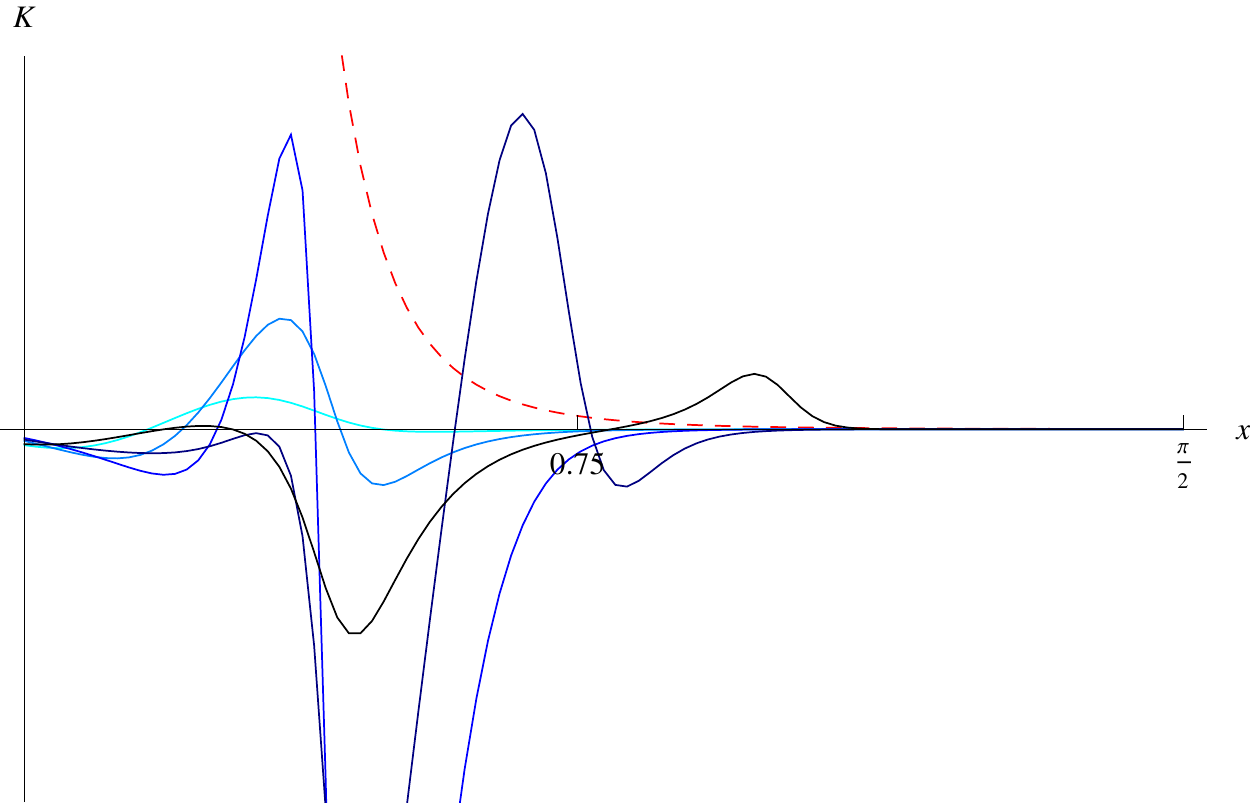}
  \caption{Plots of the Kretschmann scalar for the oscillating string along the $x$ axis for $0 < x < L/4$, where $L = 2 \pi$. The red dashed line indicates the Kretschmann scalar for a Schwarszchild black hole of equivalent mass. Plots are given at five different times, with a tenth of a period between them. For reference, the string is contained in the bounds $-0.75 \le x \le 0.75$. It is evident that there is little curvature in the inner portion of the loop.}
  \label{fig:kretschmann}
\end{figure}

The time-averaged metric produced by the string yields a Schwarzschild-like metric when far from the string, as shown in Eq. \eqref{eq:averagefield}. However, for radius $r < L$, this approximation breaks down, and the dynamics become far more complicated. To demonstrate how the inner portion of the string spacetime behaves in terms of curvature, we compute the Kretschmann scalar $R^{\mu \nu \gamma \delta} R_{\mu \nu \gamma \delta}$ along the positive $x$ axis from the origin at five different times in Figure \ref{fig:kretschmann}. We also plot the Kretschmann scalar $K = 48 G^2 M^2 / c^4 r^6$ for a Schwarzschild black hole of mass $\mu L$ for comparison. It is readily seen that the behavior of geodesics near the string in this spacetime is going to be far more complicated than for either black holes or the infinite straight string.

\section{String Microlensing} \label{sec:results}

In this section, we present the first demonstration of cosmic string loop microlensing in the form of two numerical realizations of the phenomena. We present the geometry of each situation, discuss features evident in the simulations, and quantitatively compare phenomena with theoretical predictions from the infinite straight string.

\subsection{First Demonstration of Microlensing}

\begin{figure}[t!]
  \centering
  \includegraphics[width=\columnwidth]{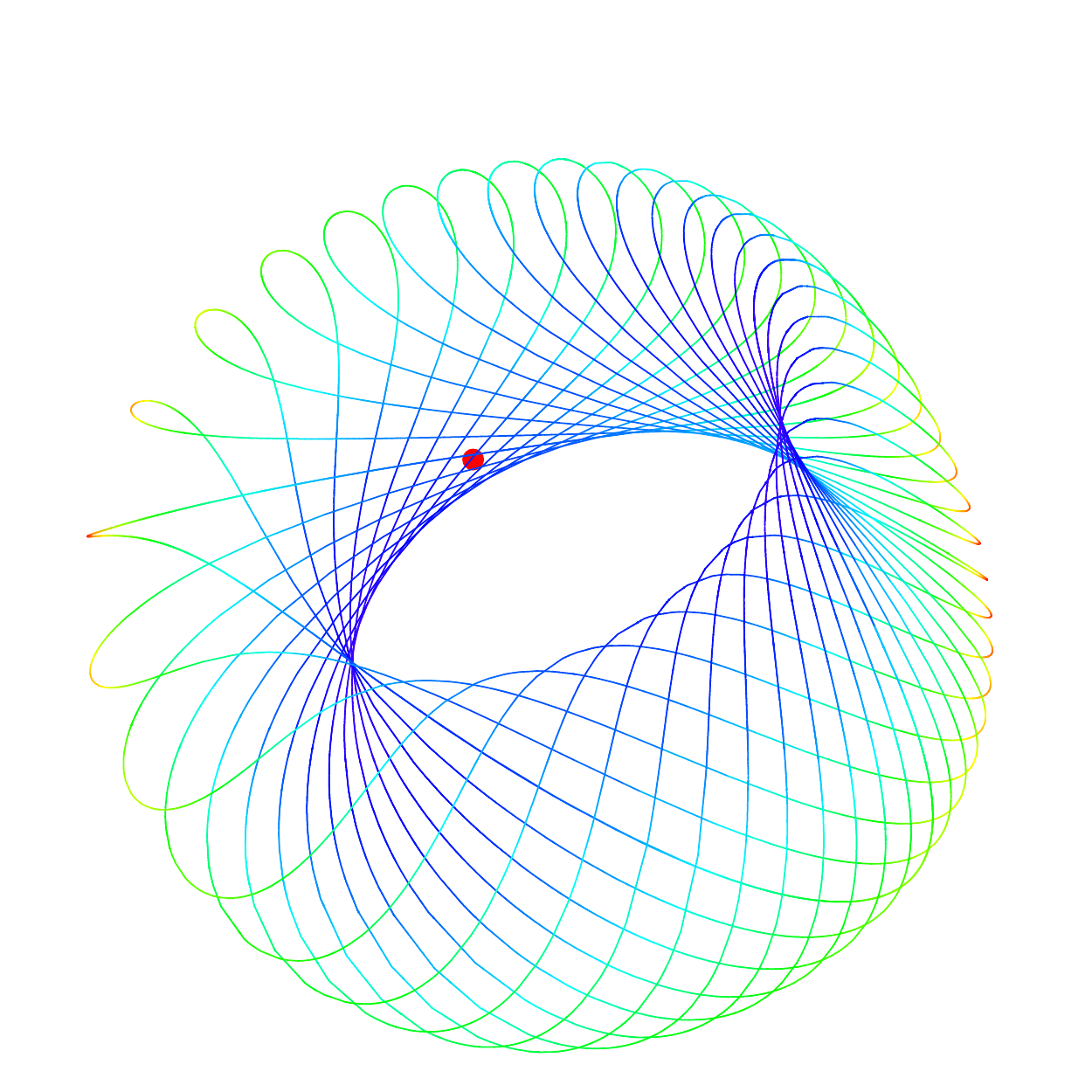}
  \caption{Plot of the retarded image of a loop throughout its period (shown in 15 steps). The red dot is the position of the source, and the loop is shown from the perspective of the observer. The colour scheme is as previously, with red indicating string velocities close to the speed of light, and blue indicating slower velocities, typically $\sim 0.1c$.}
  \label{fig:retardedloop}
\end{figure}

For our first demonstration of microlensing, we chose a simple geometry where we strongly suspected that microlensing should occur. We chose source and observer positions on opposite sides of the string such that the Euclidean path between the two points intersected the volume swept out by the string in a simple manner. This geometry is detailed in Figure \ref{fig:retardedloop}.
It is straightforward to see that the Euclidean line joining the source and observer will lie outside the loop for part of the period, and inside for the rest. This line crosses the string twice, once for a horizontal segment of string, and once for a diagonal section of string. Note that this line is nowhere near the cusps, and also avoids regions where multiple close encounters with the string might occur. The string configuration is as described in Section \ref{sec:stringconfig}, the observer is located at $(1,-4,-3)$, and the source is located at $(-1.26, 3.73, 3.3)$. In order to make the microlensing solutions as pronounced as possible, we used $G \mu = 10^{-3}$.

Using the computational method described in Section \ref{sec:computation}, we relaxed Euclidean geodesics to obtain geodesics connecting the source and observer, with arrival times spaced throughout a period. This alone could not identify microlensing solutions, as it only produced one geodesic arriving at any given time. From plots of the distance of closest approach as a function of arrival time, it is straightforward to identify when the solutions jump from geodesics passing outside the loop to geodesics passing inside the loop (although which is which requires 3D visualization tools). Solutions on either side of this jump could then be used as initial guesses from which to step forwards or backwards in time appropriately, iteratively extending the solutions until they hit the string (or in practice, got too close to the string to compute further data points in a reasonable amount of time).

\begin{figure}[t]
  \centering
  \includegraphics[width=\columnwidth]{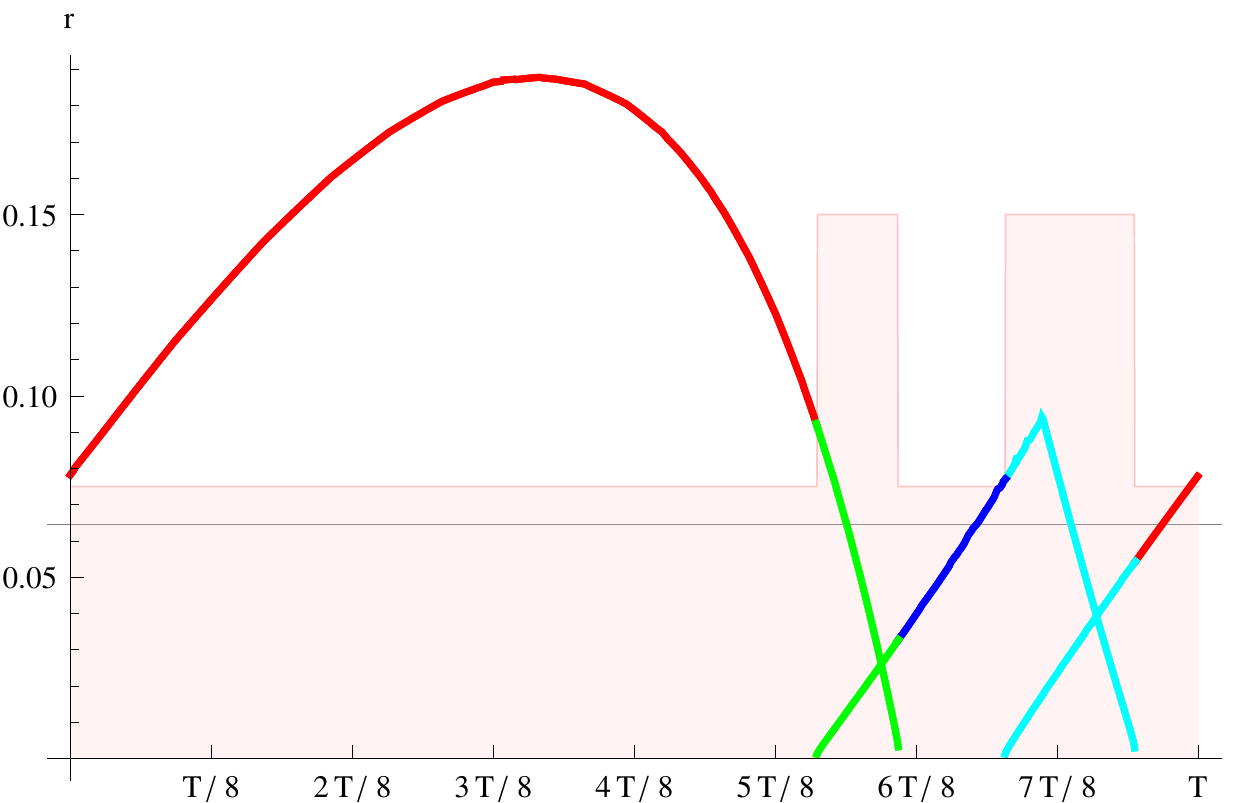}
  \caption{Plot of distance of closest approach between a geodesic and the string loop as a function of time of arrival at the observer. The distance is a Euclidean distance, and measured in the same distance units as the string (recall that we use $L = 2 \pi$). The period for this loop is $T = L/4$. The red shaded background is indicative of flux from the source across the period, in arbitrary units.}
  \label{fig:lens1dist}
\end{figure}

The results are plotted in Figure \ref{fig:lens1dist}, which shows the distance of closest approach of the geodesics arriving at different points in time. The existence of multiple solutions for some points in time is evident: we have identified microlensing solutions. The red part of the curve corresponds to geodesics lying outside the string, while the dark blue curve corresponds to geodesics passing through the string. At points in time where multiple solutions are present, the curves are coloured differently to highlight this. The black line is drawn at the theoretical estimate of the distance at which microlensing solutions exist for an infinite straight string (see Section \ref{sec:basics}). We see that with an $O(1)$ correction factor, this estimate is correct, although it is an underestimate at the beginning of the microlensing event, and an overestimate at the end. One important observation from this plot is that microlensing events begin and end when one geodesic has an impact parameter of zero.

Another quantity that can be extracted from this plot is the duration of each microlensing event, which will be important for rate calculations and potentially observing such events. The duration of these events are about 8\% and 12\% of the period. Given the estimate of Eq. \ref{eq:timing}, these durations are approximately double (triple) the estimated minimum duration for this loop. Such factors are very reasonably explained by various angles and velocity-dependent coefficients.

The jagged peak near $t = 7T/8$ is explained by dividing the string loop into two halves. To the left of this peak, one half is closer to the geodesic, but is moving away from it. On the other side, the other half of the string is closer to the geodesic, and continues to move closer. At the discontinuity in the derivative, the two segments of the loop are equidistant from the geodesic. This suggests that whenever the velocity of the string is towards a geodesic, the slope of the distance of closest approach is negative, while the opposite holds when the string is moving away from the geodesic. This fact is important when considering time delay and redshift effects below.

We investigated the linear approximation by looking at the metric perturbation at the distance of closest approach for each geodesic. We computed the maximum deviation of any component from the Minkowski metric as an indication of the size of the perturbation. The typical values we found were $h \sim 0.01$, spiking up to a maximum of $h \sim 0.07$ as the distance of closest approach became small. When applying Eq. \eqref{eq:happrox} with $r_{min} \sim 0.03$ and $\alpha \sim 0.2$, corresponding to the middle of the microlensing event, we obtain $h \sim 0.03$, and so these values are as expected. This confirms that the linear approximation works well, and shows that it is sufficient for investigations of microlensing.

\begin{figure}[t]
  \centering
  \includegraphics[width=\columnwidth]{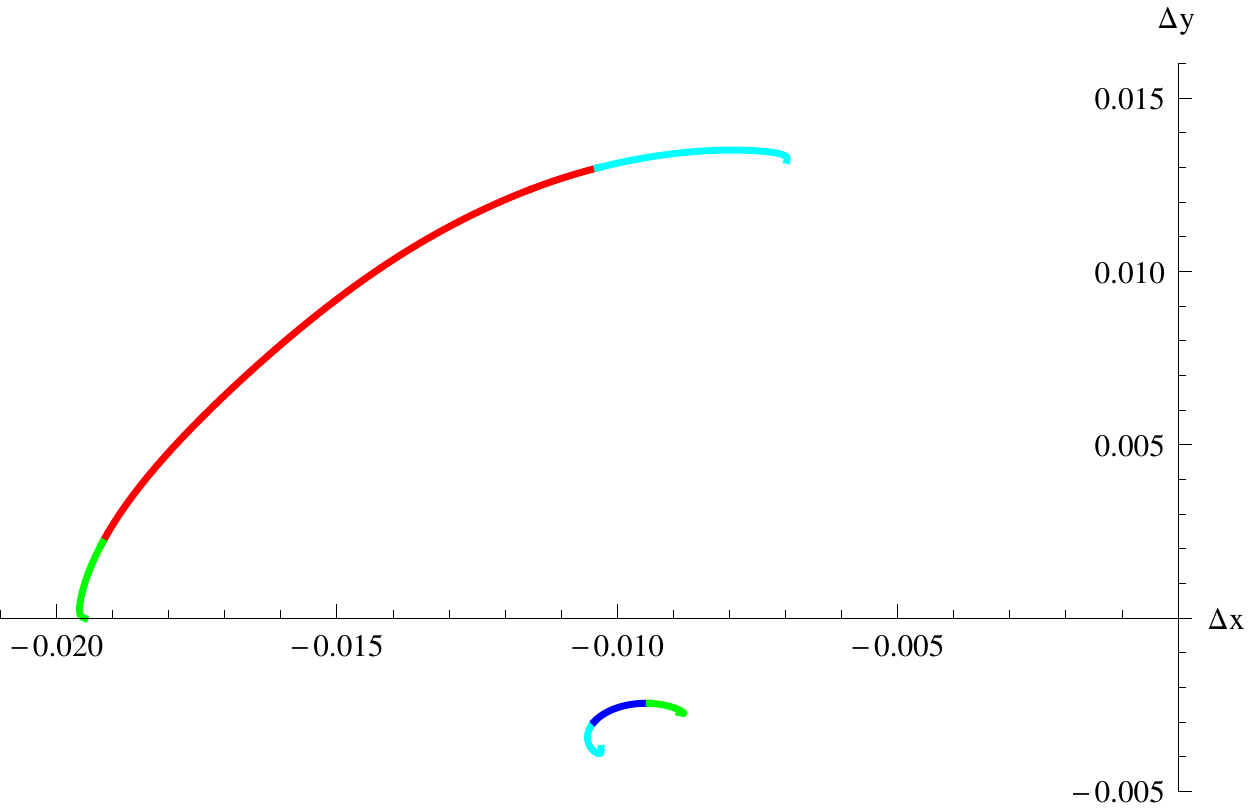}
  \caption{Plot of the angle (in radians) at which the geodesics arrive at the observer plotted in an $x-y$ plane, where the $-z$ direction points directly at the source. The colour scheme is chosen identically to the previous plot.}
  \label{fig:lens1angles}
\end{figure}

In Figure \ref{fig:lens1angles}, we plot the angle of arrival of the geodesics, choosing an arbitrary orientation. This corresponds to the path that the position of the source would track in the sky. The colours in this plot are chosen to be the same as in the previous plot, so that different curves can be matched; the locations of the source where microlensing events are occurring are evident. The origin in this plot represents the Euclidean line between the source and the observer. For small deflection angles, the difference between the two separate curves is approximately the deflection angle caused by traversing the opposite side of the string, which is roughly $0.01-0.015$ radians. As discussed in Section \ref{sec:basics}, the deflection from an infinite straight string is $8 \pi G \mu \sin \theta d_2/(d_1+d_2) \sim 0.012$ radians in this case, where we assume $\theta \sim \pi/2$. We see that with $O(20\%)$ correction factors, this deflection angle is validated by the microlensing solutions found. It is interesting to note that at the ends of both curves, where the geodesic is becoming very strained and passing very close to the string, the angle of arrival takes a sharp hook in a new direction. Closer inspection of this hook shows that it points directly towards the location of the microlensed solution. Another interesting point to note is the ``Newtonian'' deflection angle, caused simply by the mass of the string. While it varies depending on how the string is configured at any given point in time, it is of the same order as the deflection from the deficit angle.

\begin{figure}[t]
  \centering
  \includegraphics[width=\columnwidth]{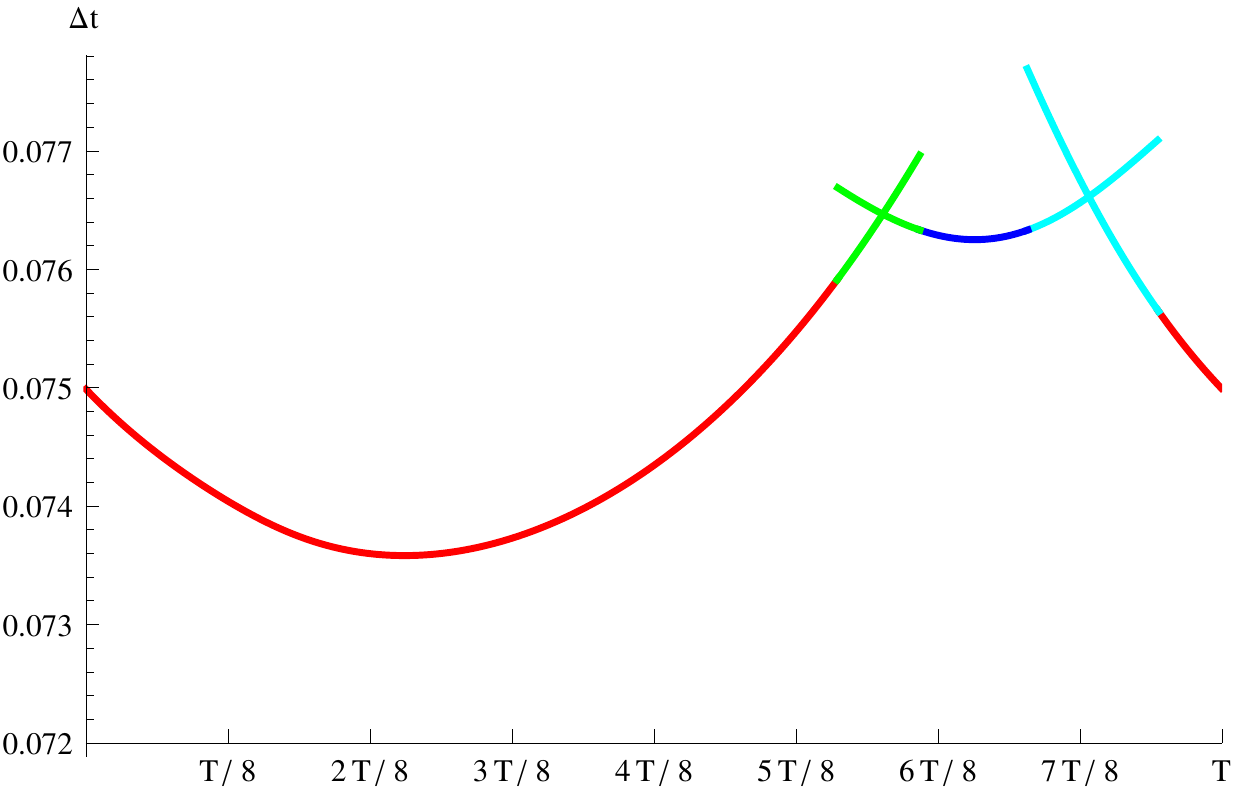}
  \includegraphics[width=\columnwidth]{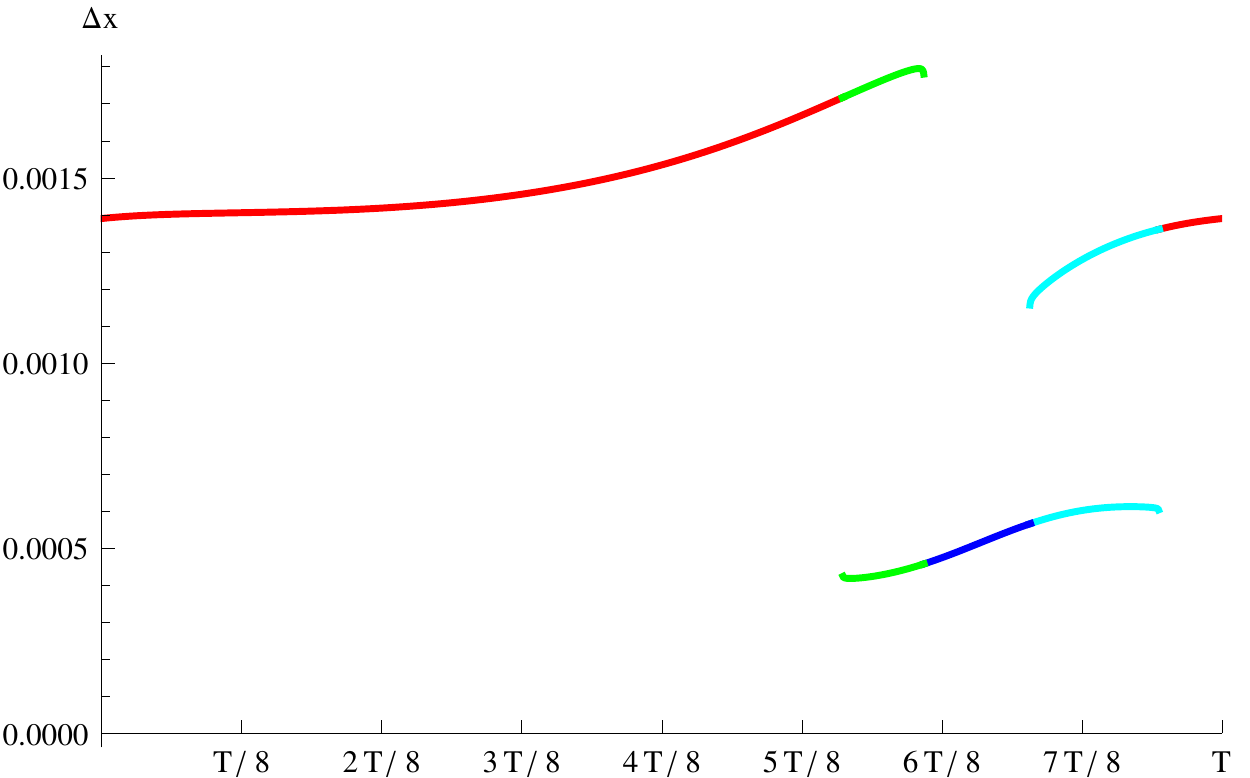}
  \caption{Above: plot of the time delay of geodesics as a function of arrival time, compared to the Euclidean travel time.\\
  Below: plot of the increased Euclidean distance through geodesic bending as a function of arrival time.}
  \label{fig:lens1time}
\end{figure}

We can also investigate the time of flight of the geodesics. These are plotted in Figure \ref{fig:lens1time}, where the $y$ axis plots the time delay compared to the Euclidean distance between the observer and the source. We immediately see that the time delay is always positive. The magnitude of the delay is due to a number of factors. The most obvious reason is that the geodesics are bent, and are thus travelling a greater distance. In the lower figure, we plot the Euclidean distance travelled by the geodesics (ignoring $O(h)$ corrections). This plot shows that the extra distance traversed because of the bent geodesics incurs only $1-2\%$ of the observed extra travel time. It also shows that the travel distance is roughly constant for each curve, and that geodesics passing through the loop have a slightly shorter journey. The dominant contribution to the time delay is from the Shapiro time delay \cite{Shapiro1964}. The time delay from this effect is proportional to $G \mu L \sim 6 \times 10^{-3}$ with a coefficient depending upon the geometry, which could easily be $O(10)$. When the geodesic is far from the string, we see that the travel time is less. Similarly, the curve describing geodesics that travel through the string loop is significantly more delayed. We attribute this effect to the coefficient describing the Shapiro time delay increasing as the geodesics moves deeper into the potential well of the string. As the geodesics approach the string more closely, they move into stronger wells, explaining the increased time delay at the ends of the curves. Note that the intersections between microlensing solutions that occur on this plot do not occur at the same times as the intersections on the plot showing distance of closest approach. It is intriguing that these curves intersect, as it is not clear that this is required.

\begin{figure}[t]
  \centering
  \includegraphics[width=\columnwidth]{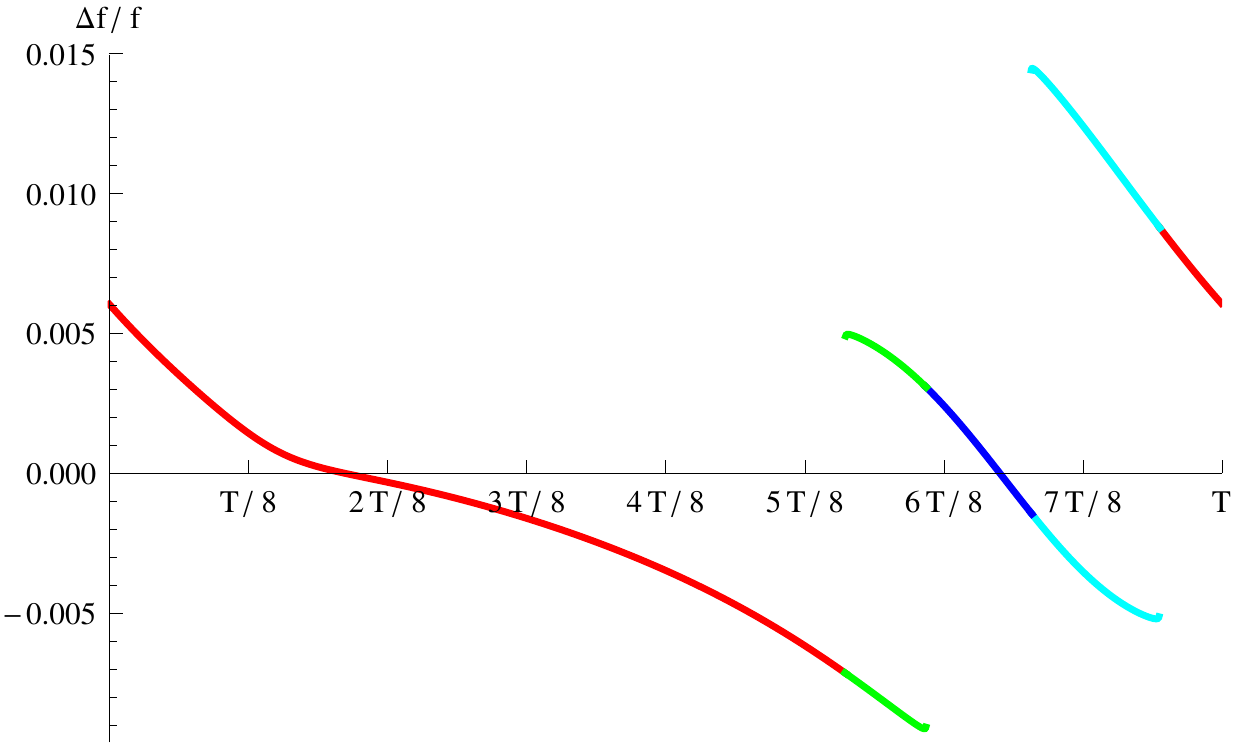}
  \caption{Plot of the redshift $(f_{obs}-f_{src})/f_{src}$ for geodesics as a function of arrival time.}
  \label{fig:lens1redshift}
\end{figure}

Finally, we can plot the redshift of the geodesics $(f_{obs} - f_{src}) / f_{src}$, as shown in Figure \ref{fig:lens1redshift}. A general trend is evident: when the string is moving away from the geodesic, photons become blueshifted, while when the string is moving towards the geodesic, the photons become redshifted. The size of the shift red/blueshift varies between $\sim \pm 4 \pi G \mu$, in good agreement with the prediction from the Kaiser-Stebbins effect \cite{Kaiser:1984iv} described in Section \ref{sec:basics}.

\begin{figure}[t]
  \centering
  \includegraphics[width=\columnwidth]{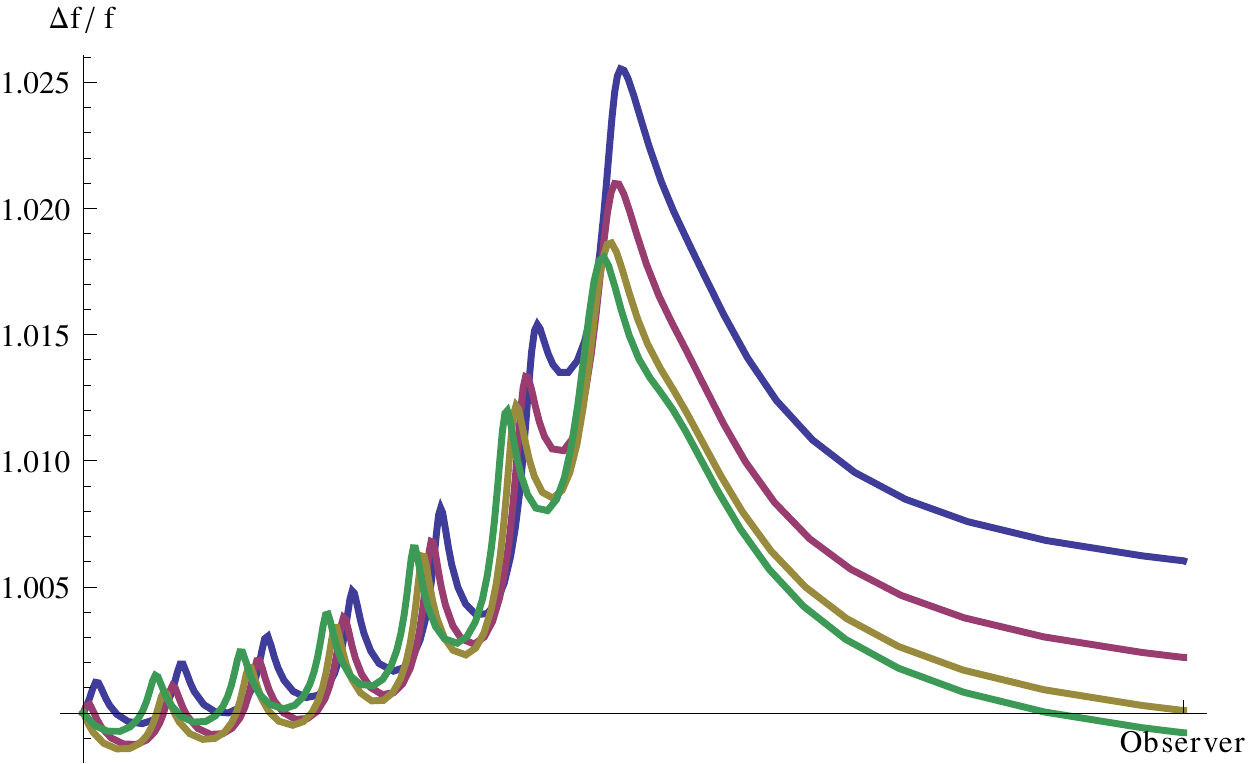}
  \caption{Plot of redshift in a geodesic as a function of position. The $x$-axis ranges from the source to the observer, while the $y$-axis shows the redshift. Four curves are plotted, with a time difference of a tenth of a period between them. These curves are all from geodesics that pass outside the loop.}
  \label{fig:lens1geodesic}
\end{figure}

The final plot that we include in Figure \ref{fig:lens1geodesic} is the redshift along a geodesic as a function of position. We see that as a photon leaves the source and travels towards the string, it encounters waves of gravitational radiation. In the middle of the loop, the photon is blueshifted by the potential well, and as it leaves, it ``surfs'' a wave at a redshift that decays as $1/r$ (the decay of a gravitational wave from a cosmic string loop). The final redshift is somewhat dependent upon whether the photon was emitted at a peak or a trough of the gravitational waves. These observations coincide with observations about the computational time necessary for integrating geodesics: the path of the photon approaching the loop was always much more computationally intensive than the path of the photon leaving the loop.

One thing that was not done in this analysis was to transform the source and observer's coordinates into an orthonormal basis in order to compute the deflection angles and redshift, justified on the basis that the source and observer are sufficiently far from the string that the use of the Minkowski metric is a good approximation. For most phenomena, this would only give rise to an $O(h)$ correction to $O(h)$ effects, and as the distance from the observer to the string grows, the effects can only shrink. However, the effects on redshift may be more pronounced, because of the dependence of whether the photon was emitted during a peak or a trough of the gravitational radiation.

\begin{figure*}[p]
  \centering
  \includegraphics[width=0.8\textwidth]{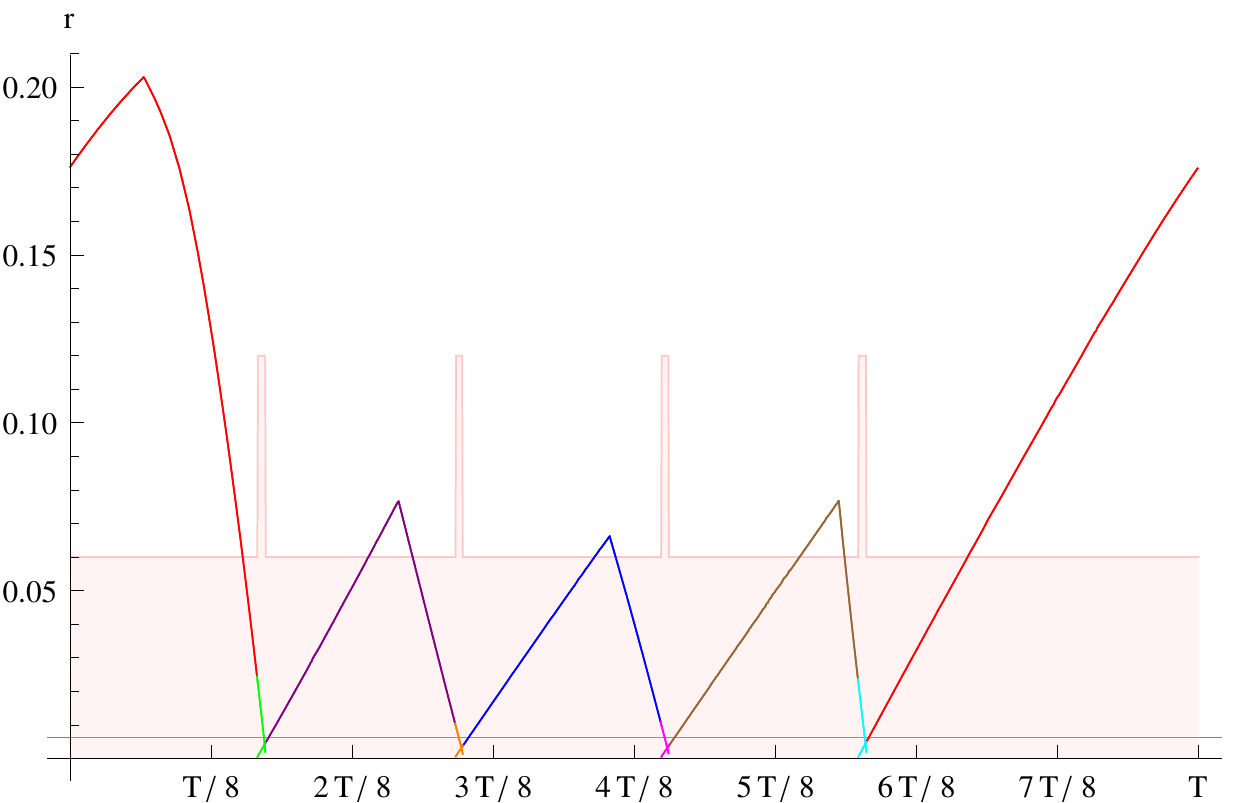}
  \includegraphics[width=0.8\textwidth]{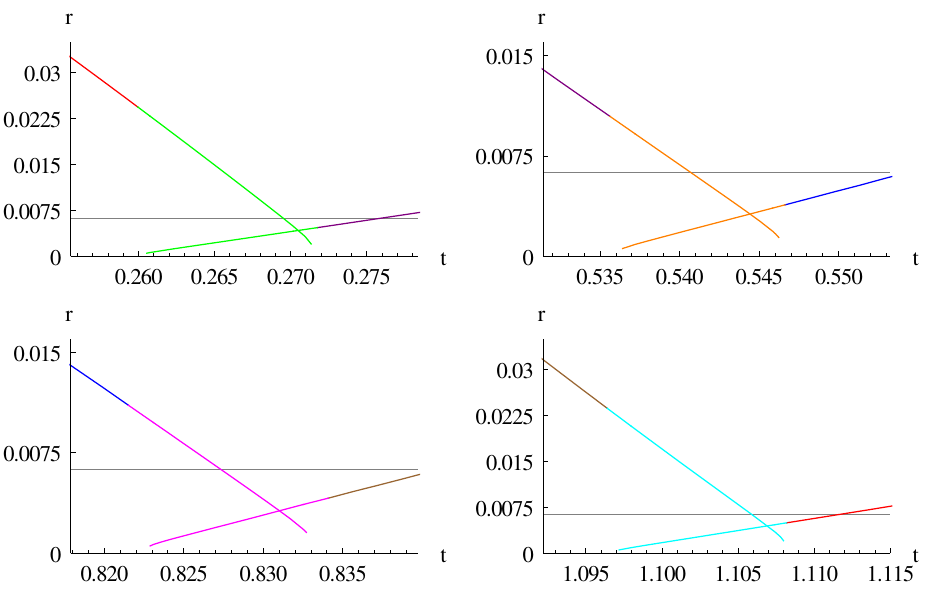}
  \caption{Plot of the distance of closest approach between a geodesic and the string loop as a function of arrival time. The red shaded background is indicative of flux from the source across the period, in arbitrary units. The lower plot shows a zoomed in version of the four microlensing events. Recall that $T = L/4 = \pi/2$ in our units. Again, the four individual curves are colour-coded to identify which curve is which on different plots, and the colour also changes to reflect the existence of multiple images at that time.}
  \label{fig:lens2dist}
\end{figure*}

\subsection{Generic Lensing}

Our first example of cosmic string microlensing demonstrated all of the expected effects predicted analytically from analysing the infinite straight string system. However, it represented a carefully chosen system where we expected microlensing to occur, using geodesics that grazed the side of the loop appropriately, and was also performed at a very strong string tension. For our second analysis of string microlensing, we decided to choose an arbitrary direction (distinct from the previous direction), and place the source and observer on opposite sides of the loop in that direction. The observer was placed along the $z$-axis at $(0,0,5)$, and the source at $(0,0,-5)$. Euclidean geodesics connecting the two thus run directly through the center of the string loop. We also decreased the string tension by an order of magnitude to $10^{-4}$. This configuration demonstrated all of the effects we previously identified, and also contained some unexpected surprises.

\begin{figure}[tb!]
  \centering
  \includegraphics[width=\columnwidth]{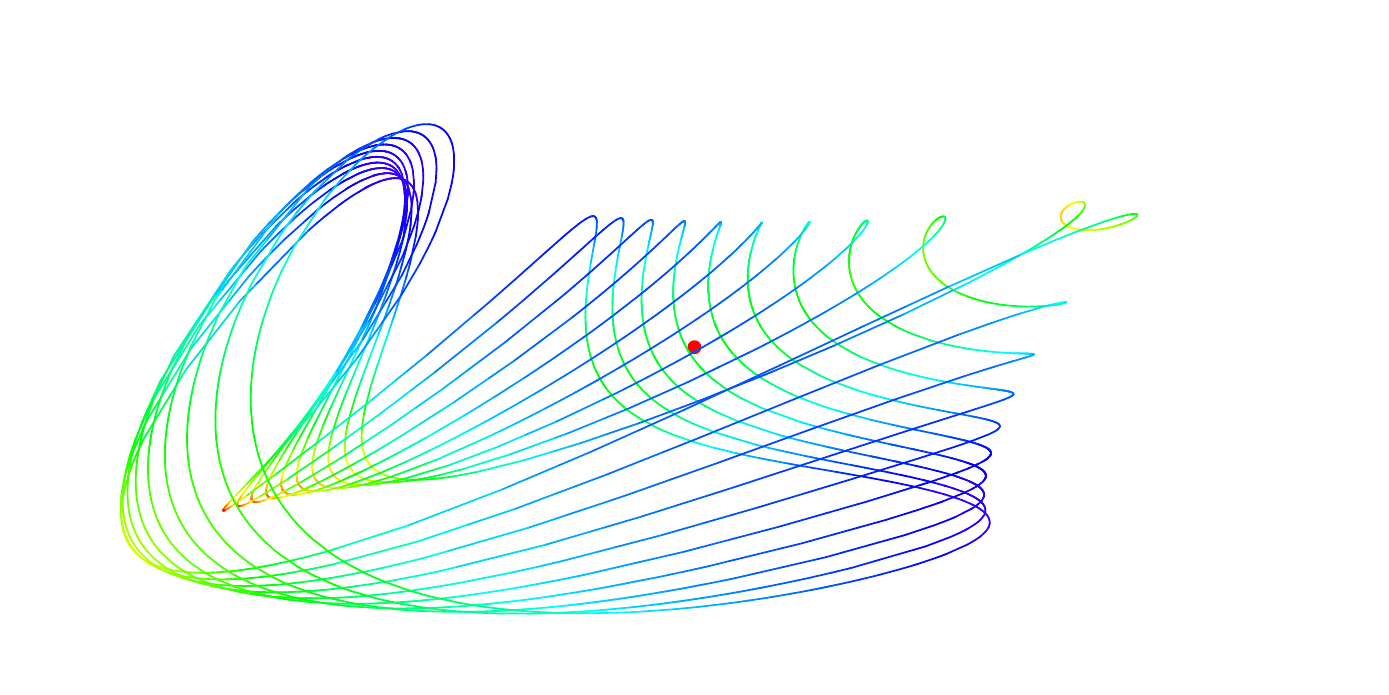}
  \includegraphics[width=\columnwidth]{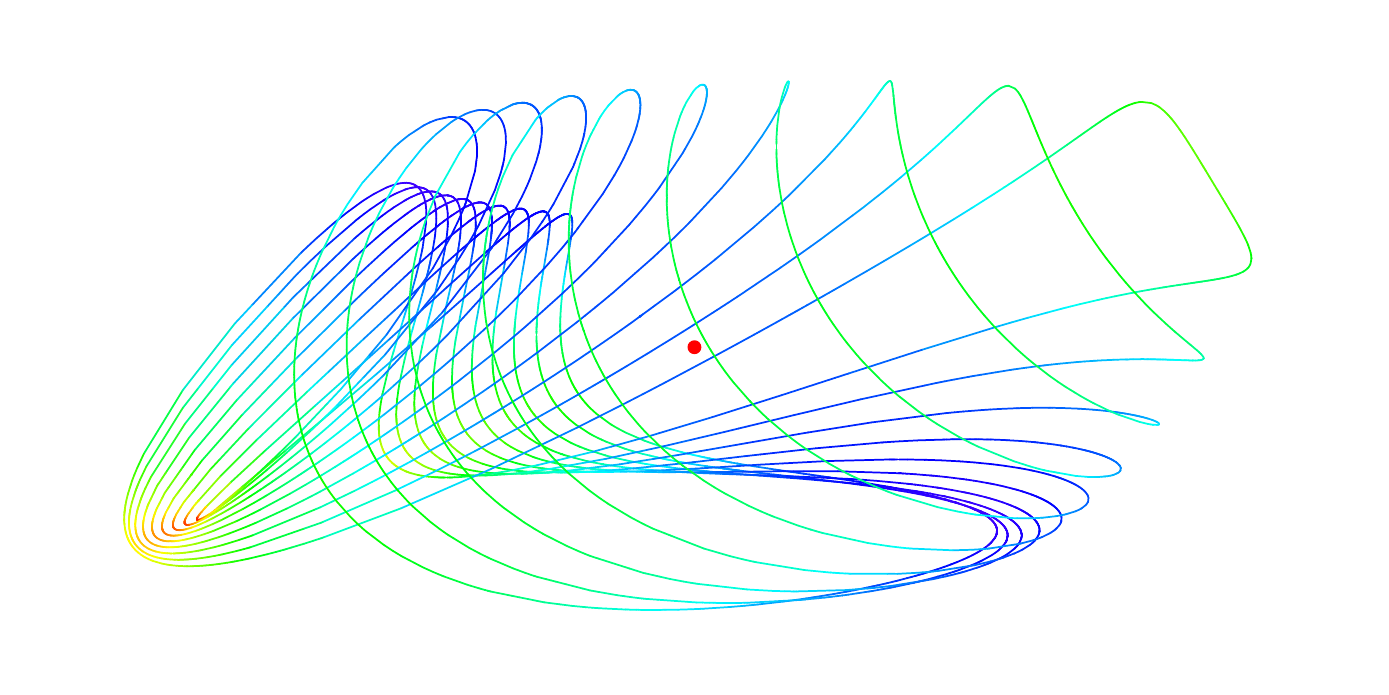}
  \caption{Plots of the retarded image of a loop throughout its period, shown in two halves (10 steps each). The red dot is the position of the source, and the loop is shown from the perspective of the observer. In the top image, the loop segment that crosses the line between the observer and the source is closer to the source than the observer; the reverse is true for the bottom image.}
  \label{fig:lens2retarded}
\end{figure}

The retarded image of the string is shown in Figures \ref{fig:lens2retarded}. We see that in each half of the period, we can expect two microlensing events to occur, as the string loop crosses the Euclidean geodesic twice in each image, once diagonally to the right, and once diagonally to the left. Note that the cusp on the bottom left, visible in red in the top plot, is visible for a number of images. Conversely, the cusp on the right, barely visible at all in the top right corner of the bottom plot, is only visible for a single image. The reason behind this is that cusp in the top image is moving away from the observer close to the speed of light, and so the retarded image sees it at a number of instants in time. The cusp in the bottom image, however, is moving towards the observer close to the speed of light, and so the window in which the retarded image sees the cusp is significantly shortened.

\begin{figure*}[p]
  \centering
  \includegraphics[width=0.8\textwidth]{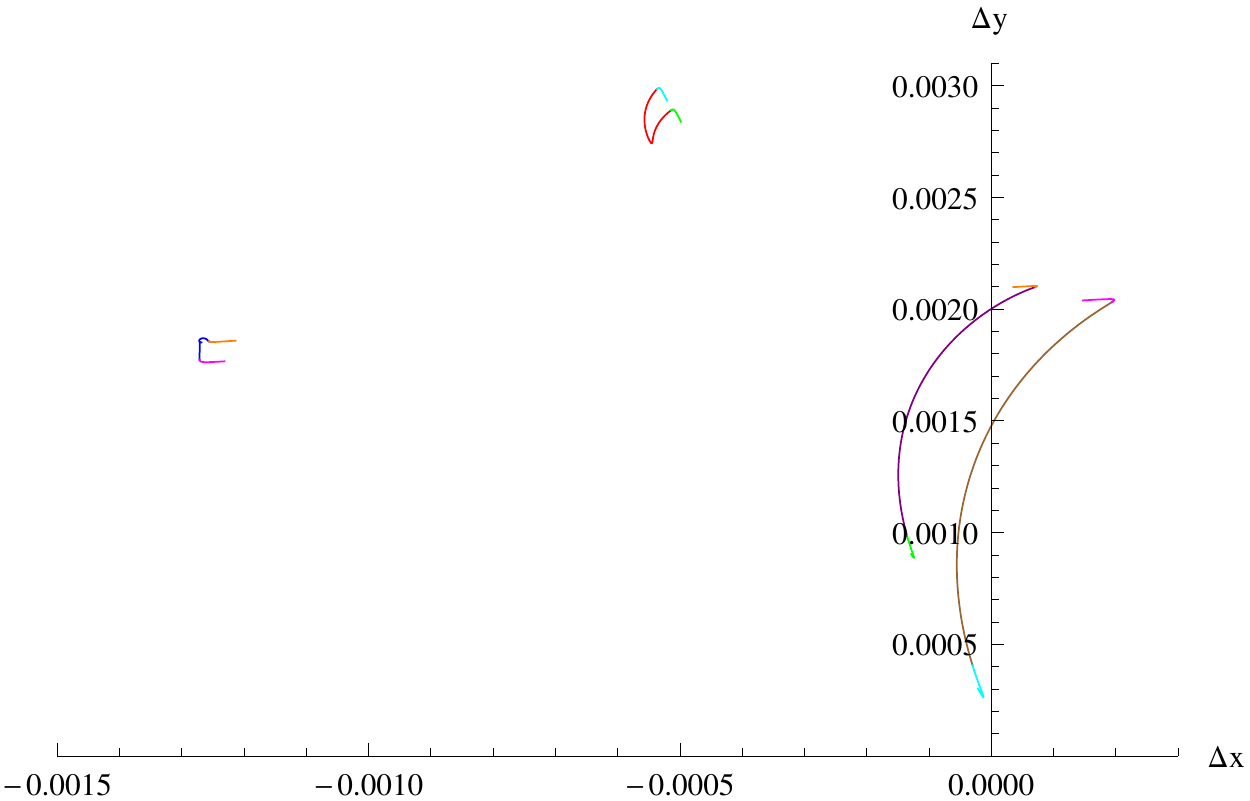}
  \includegraphics[width=0.8\textwidth]{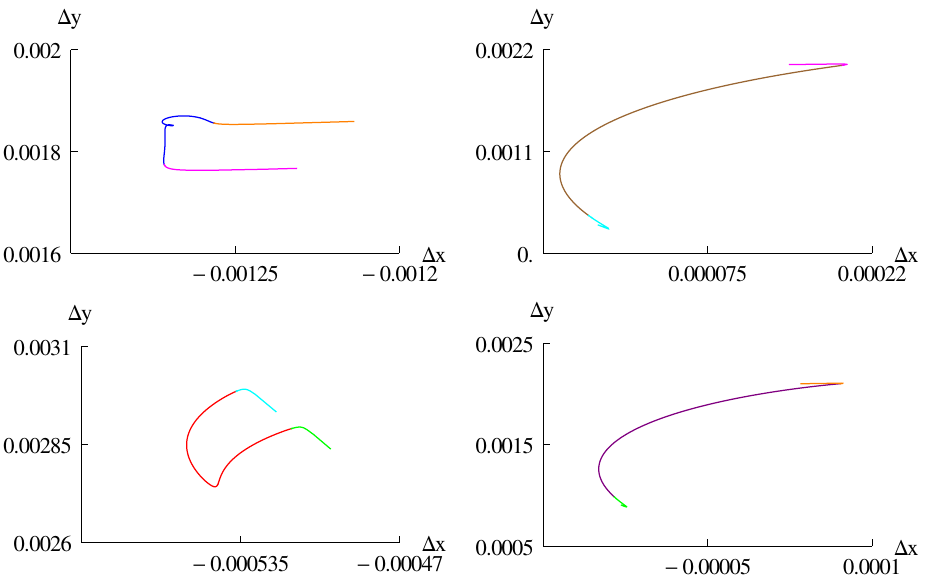}
  \caption{Plot of the angle (in radians) at which the geodesics arrive at the observer plotted in an $x-y$ plane, where the $-z$ direction points directly at the source. The bottom four plots show zoomed in versions of the individual curves.}
  \label{fig:lens2angles}
\end{figure*}

Our heuristic picture is confirmed in the plot of distance of closest approach as a function of arrival time, shown in Figure \ref{fig:lens2dist}. We see that indeed, four microlensing events occur. Because the string tension is an order of magnitude lower this time, the theoretical prediction at which microlensing events occur is substantially lower, and extracting data in the microlensing regime was computationally intensive, as the points needed to be an order of magnitude closer to the string than in the previous case. We see that the theoretical estimate of when microlensing should occur is again correct to within a factor of a few, but once again, provides an underestimate at the beginning of the lensing event, and an overestimate at the end. This effect is
likely related to the velocity of the nearest string segment and the angle of approach of the geodesic.

We see that there are four jagged peaks in the curves which are again explained by different segments of the string becoming closer. Looking at the zoomed-in plots of the microlensing events, note that the plots occur in pairs: the top left and bottom right plots are almost identical, while the top right and bottom left plots are also almost identical in timing, distance, and slopes for all curves. The duration of the microlensing events were, in order, approximately 0.7\%, 0.65\%, 0.65\% and 0.7\% of the total period. These are approximately twice as long as the predicted minimum duration for a microlensing event, which can again be very reasonably explained through appropriate angles and velocity-dependent factors. Although we present only approximate figures, reflecting uncertainty in the precise timing, we note that the duration of the first and final events were numerically identical to four significant figures, as were the second and third events, further suggesting some symmetry between the events.

We conjecture that this symmetry lies in the Burden loop, which oscillates in a symmetric fashion. Looking at Figure \ref{fig:lens2retarded}, we see that the first two microlensing events occur when the geodesics pass near segments of the string closer to the source, while the other two occur for segments of the string nearer the observer. However, because of the loop's symmetry, the segments of string at the front and the back are moving at the same velocity, which leads to the similarities in the microlensing events. The symmetry cannot be exact, however, as the geodesic's behavior before and after the close encounter with the string is different in the different cases (in one case, it's entering the middle of the string, and in the other, it's leaving).
The duration of a microlensing event is primarily determined through the details of the close encounter with the string, while the complicated dynamics outside this regime have little impact upon this result.

The linear approximation at the point of closest approach was once again investigated for each geodesic. In this case, the typical size of the metric perturbation was $h \sim 10^{-3}$, spiking up to $h \sim 0.014$ for the very closest geodesics, again showing that the entire simulation was safely in the linear regime. Compared to our previous simulation, the decreased string tension is slightly offset by the nearer distance of closest approach. The analytic estimate from Eq. \eqref{eq:happrox} with $\alpha \sim 0.2$ and $r_{min} \sim 0.075$ yields $h \sim 0.004$, which is well within the expected regime.

The next plots, Figure \ref{fig:lens2angles}, show the angular position of the source on the sky, as seen by the observer. Again, the rotation is chosen arbitrarily. As expected, there are four separate curves in the sky. Contrary to the previous example, where the outside image moved a substantial angle through the sky, here, the outside images are the red and blue curves, which have much smaller footprints on the plot. The images that arise from threading the loop move a more significant angle through the sky. The similarities in these two curves are ascribed to the previously noted symmetry. Again, when microlensing solutions exist, the angle between them is roughly $4 \pi G \mu$, in agreement with the theoretical prediction.

There are a number of interesting features to this plot. When microlensing solutions exist, the curves traced in the sky change direction abruptly. Furthermore, they move in the direction of where their partner solution lies. Looking at the previous plots in Figure \ref{fig:lens2dist}, we see that the microlensing solutions only occupy a small portion of the time along the curves, but the angular distance travelled in that time can be significant, especially in the outside curves.
A subtle feature, the small blip in the blue curve in the top left plot, turns
out be quite interesting and we will return to this in a moment.

\begin{figure}[tb!]
  \centering
  \includegraphics[width=\columnwidth]{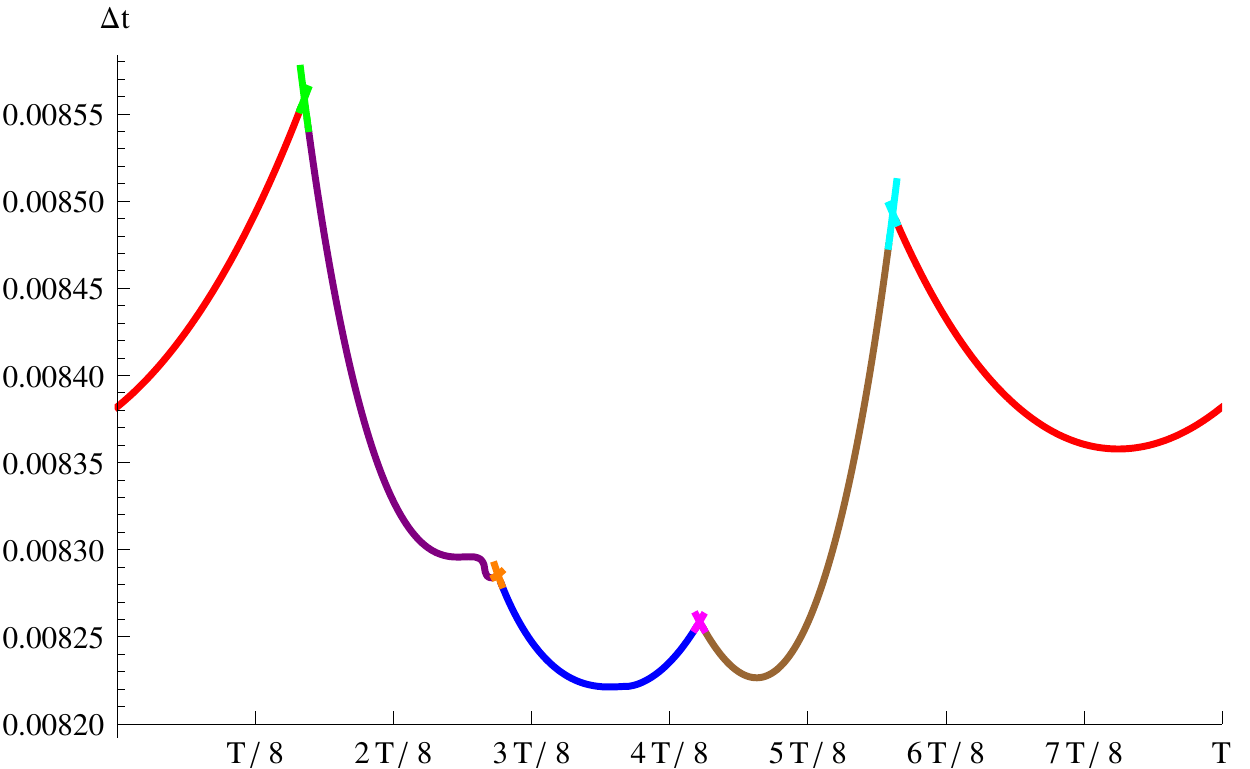}
  \includegraphics[width=\columnwidth]{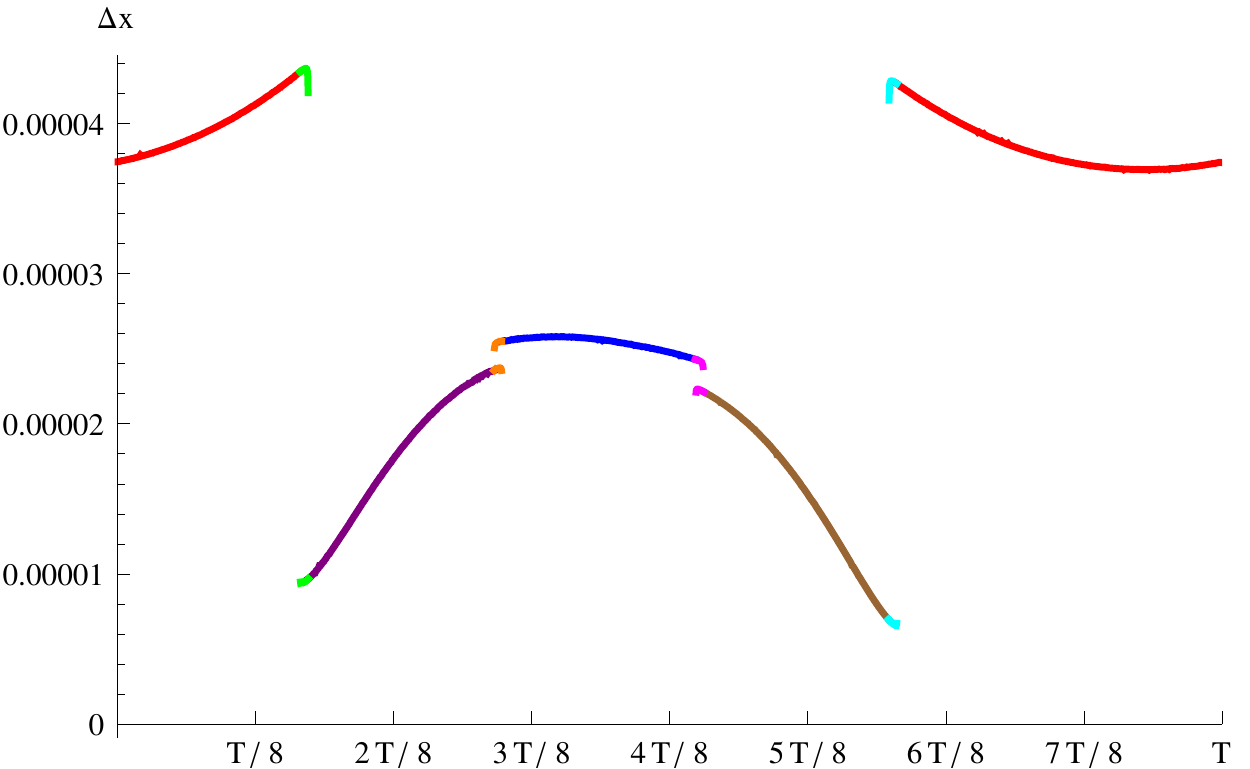}
  \caption{Above: plot of the time delay of geodesics as a function of arrival time, compared to the Euclidean travel time.\\
  Below: plot of the increased Euclidean distance through geodesic bending as a function of arrival time.}
  \label{fig:lens2time}
\end{figure}
We next look at the time delay and Euclidean distance increase, as shown in Figure \ref{fig:lens2time}. We continue to see the general trends previously identified. The distance increase cannot account for the time delay, so the bulk of it must come from the Shapiro time delay. The time delays between different curves again cross somewhere during the microlensing events.
The Euclidean distance increase is greatest for geodesics that do not pass through the loop, showing that they are more bent.
A subtle feature of note in these plots is in the time delay of the purple curve near $t\sim 5T/16$, which contains an elbow. Again, we return to this feature later.

\begin{figure}[t]
  \centering
  \includegraphics[width=\columnwidth]{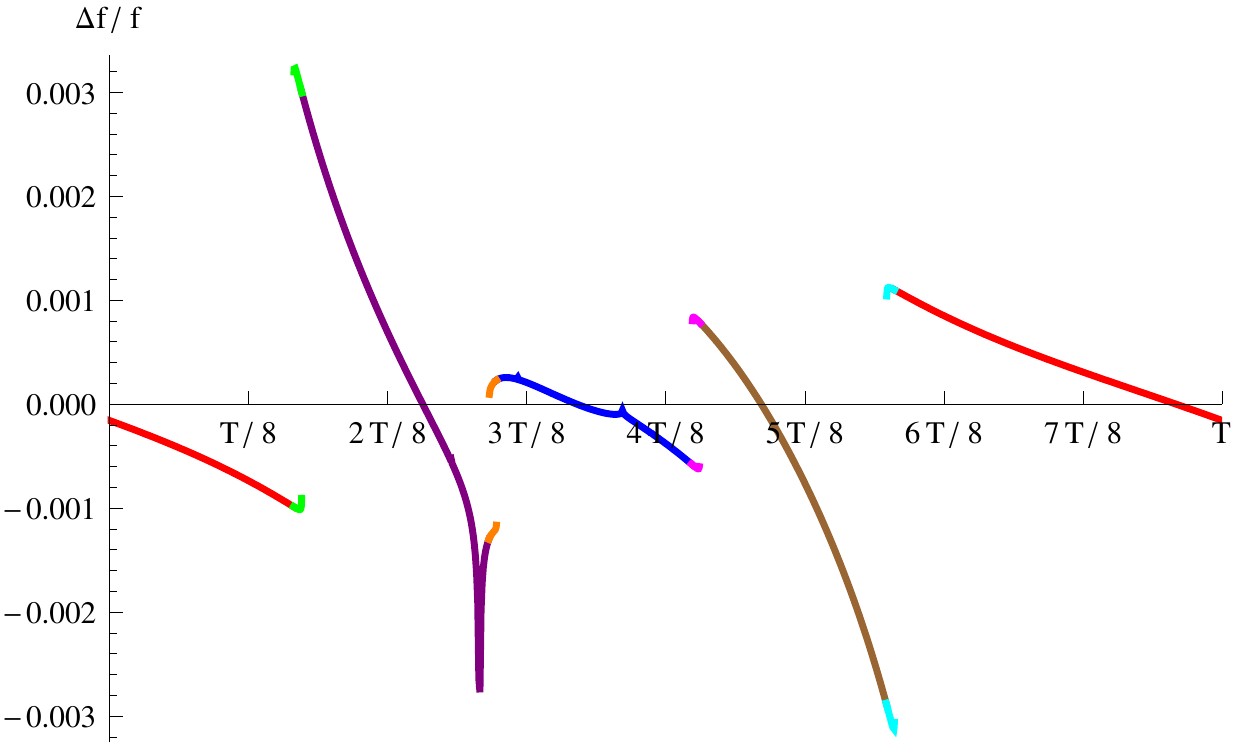}
  \includegraphics[width=\columnwidth]{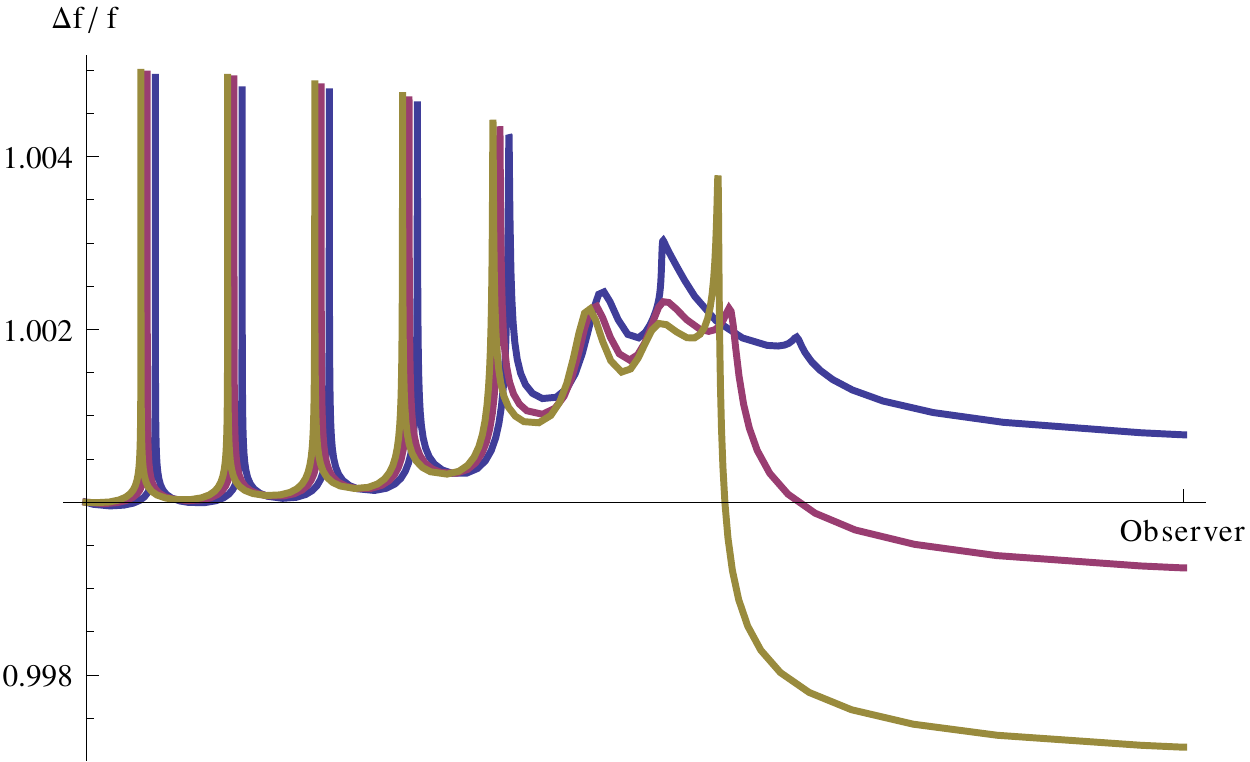}
  \caption{Above: plot of the redshift $(f_{obs}-f_{src})/f_{src}$ for geodesics as a function of arrival time.\\
  Below: plot of redshift in a geodesic as a function of position. The $x$-axis ranges from the source to the observer, while the $y$-axis shows the redshift. Three curves are plotted, at times corresponding to the beginning (top, blue), middle (middle, purple) and end (bottom, gold) of the brown curve (near $5T/8$). These curves are all from geodesics that pass inside the loop.}
  \label{fig:lens2redshift}
\end{figure}
We now turn to the redshift of the geodesics in Figure \ref{fig:lens2redshift}. The general trend described by the Kaiser-Stebbins effect continues to hold here: photons are blueshifted when the nearest string loop segment is moving away from the geodesic, and redshifted when it is moving towards the geodesic. We also see some interesting new features. In the top graph, we see a spike in the purple graph around the same position as the previously noted elbow and
a blip in the blue curve, around the position it showed on the angular deflection plot, suggesting that whatever caused those features previously also impacts the redshift.

The other interesting feature to note is in the lower plot. Previously, there were troughs and crests to a photon propagating against the gravitational radiation into the string loop. Here, the redshift curves look much more like the absorption of a high-finesse cavity. We conjecture that this arises because of the direction in which the geodesics approach. Recall that the string loop is mostly in the $x$-$z$ plane, and the geodesics move mostly in the $z$ direction. This means that both the near and the far side of the loop can generate
metric peturbations at the fundamental frequency, and constructive interference gives rise to the large spikes in the redshift. These spikes are even stronger than the potential well in the middle of the loop.
We will loosely refer to these perturbations as radiation even though they occur in the near zone of the system. On the far side of the loop, the photons once again ride the gravitational radiation as it falls off as $1/r$.

\begin{figure}[t]
  \centering
  \includegraphics[width=\columnwidth]{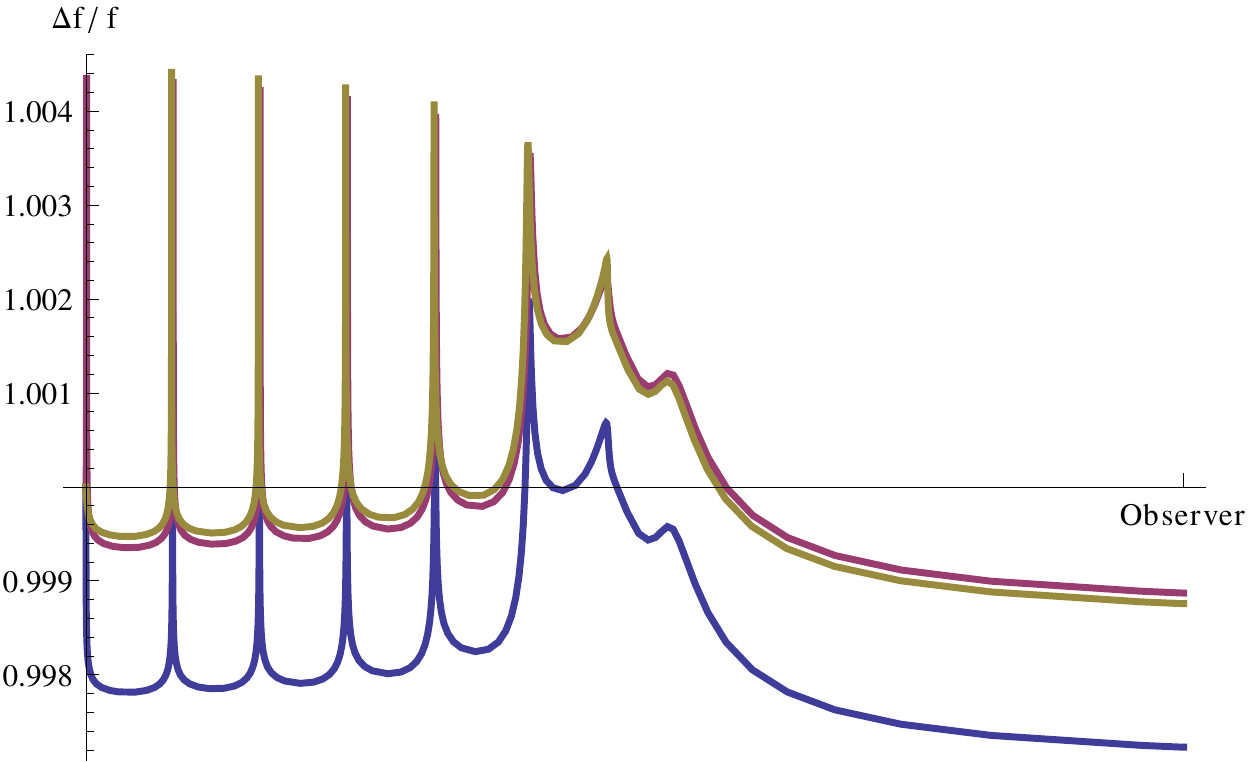}
  \caption{Plot of the redshift along a geodesic at three times: just before, during, and just after the spike in the purple curve.}
  \label{fig:lens2spike}
\end{figure}
Let us now turn to the feature in the purple curve that was clearly evident in the redshift plot and also visible in the time delay plot. The reason for this spike can be understood in terms of the crests of the gravitational radiation. Once per period, the source is going to emit radiation starting on a crest; as the photon leaves the crest, it will be redshifted down the crest. In Figure \ref{fig:lens2spike}, we see the redshift curves before, during and after the crest passes the source, which confirm this as the source of the feature. This behavior was not identified in the previous lensing simulation, as the crests of the gravitational radiation were not nearly so pronounced.

\begin{figure}[t]
  \centering
  \includegraphics[width=\columnwidth]{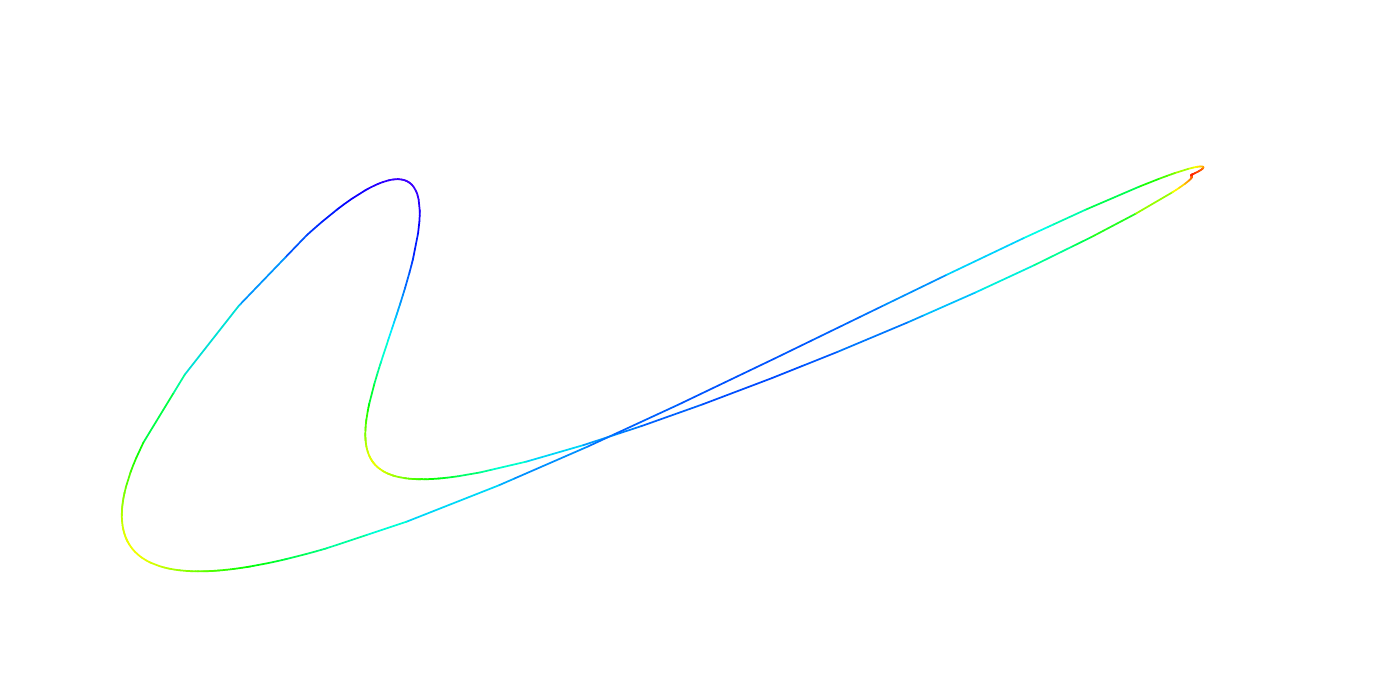}
  \caption{Retarded image of the string loop from the perspective of the observer at the time of the feature in the blue curve. The image is colour-coded by the velocity of the string, with red being close to the speed of light, and blue being much slower (around 0.1c). The formation of the cusp on the right of the curve is evident, which is moving roughly towards the observer.}
  \label{fig:lens2blueblipcusp}
\end{figure}
\begin{figure}[t]
  \centering
  \includegraphics[width=\columnwidth]{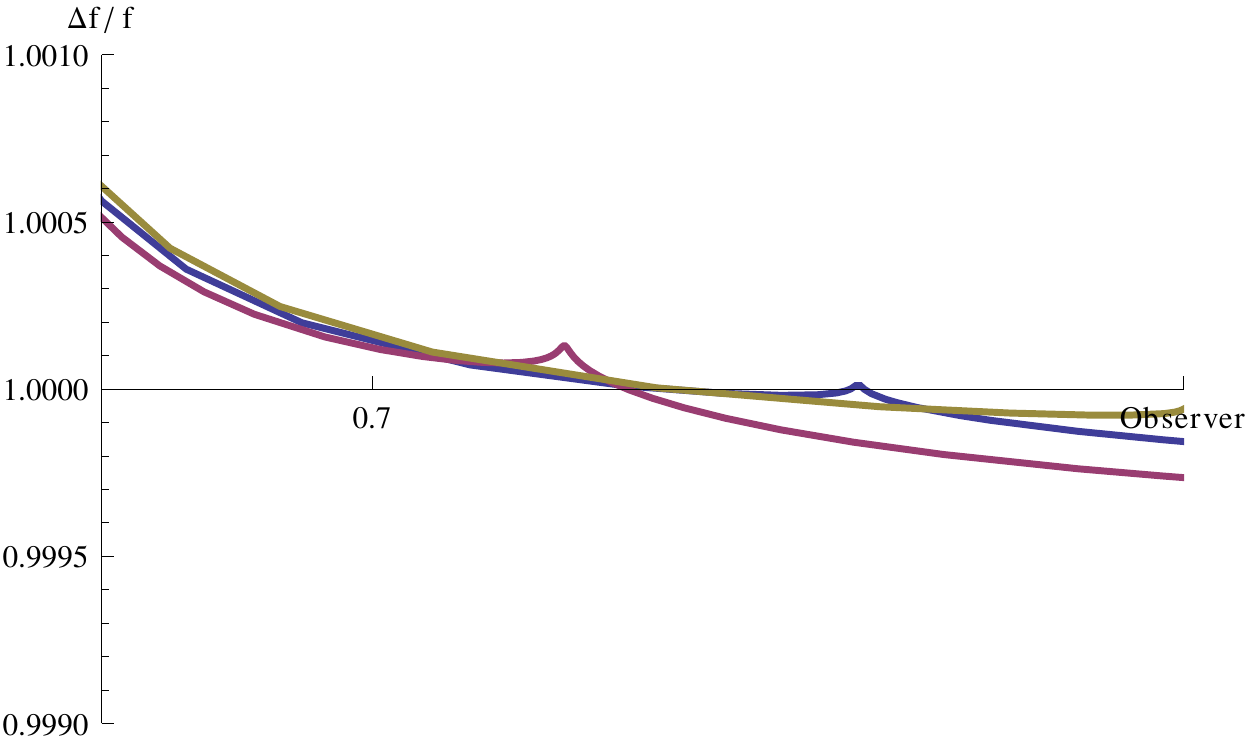}
  \caption{Plot of the redshift along a geodesic at three times: at the time of the feature in the blue curve (gold), and at two later times. This plot only shows the last 40\% of the distance from the source to the observer, in order to highlight the unexpected features of the tail end of the geodesic.}
  \label{fig:lens2blueblip}
\end{figure}
The final feature to explain is the blip in the angular deflection of the blue curve, and the accompanying blip in the redshift plots. This feature is serendipitous: it turns out that the position of the observer is within a cone of cusp radiation, as one of the cusps from the loop radiates in the $+z$ direction. The fact that the cusp radiation is responsible is corroborated by the retarded image of the string at the time of that blip, as shown in Figure \ref{fig:lens2blueblipcusp}: a small red dot to the right is moving at the speed of light; it happens to be moving roughly in the direction of the observer. We can also look at the redshift along the geodesic that arrives at the time of that blip; it shows an obvious upswing at the very end that is not expected from the $1/r$ decay of gravitational radiation as photons leave the string loop, see Figure \ref{fig:lens2blueblip}. Even more telling is the redshift along geodesics some time later, which show features suggesting that they too have been affected by a burst of radiation passing through.
We find that a null geodesic connects these events to the spacetime position of the cusp. The blue curve in Figure \ref{fig:lens2redshift} displays a corresponding blip associated with the same phenomenon. Although the cone of cusp radiation is typically thought of as being quite narrow, the opening angles required here are $\sim 0.15$ radians to affect the observer, and up to $\sim 0.3$ radians to create the feature in the purple curve in Figure \ref{fig:lens2blueblip}. A similar cone of cusp radiation will affect the source, but this signal has likely been drowned out by the dynamics as photons propagate through the inner part of the string loop.

\section{Conclusion}\label{sec:conclusion}

In this paper, we presented numerical solutions for microlensing from cosmic string loops. Two specific configurations were analyzed in detail; one was carefully constructed to yield microlensing solutions and demonstrate the robustness of our method, while the second was chosen more arbitrarily and investigated at smaller string tension. Both configurations showed clear evidence of the existence of microlensing solutions.

A number of theoretical predictions from the infinite straight string were tested in the loop configurations. First and foremost, the existence of multiple geodesics between the source and observer was verified. The theoretical prediction of the distance of closest approach at which multiple solutions occur was validated, and the deficit angle prediction for the angle of deflection between multiple solutions was also shown to hold true. The Kaiser-Stebbins effect on photon redshift was also demonstrated. Each of these effects were shown to be qualitatively true, sometimes with $O(1)$ corrections from the string dynamics.

We identified a number of features of cosmic string microlensing that are absent for the infinite straight string. The most significant of these was the contribution to the time delay arising from the Shapiro effect. Another was the magnitude of the deflection of a geodesic from the string's mass, as opposed to the deficit angle. It was found that this deflection angle is of the same order as the deficit angle. The effect of the cone of cusp radiation on photons was observed, which caused a redshift and angular deflection. This was despite not selecting geodesics that passed particularly close to a cusp. Finally, a redshift effect based on the gravitational wave phase at the source emission was observed. This was particularly evident in the second configuration we analyzed, where an unexpected resonance gave rise to strongly peaked pulses of gravitational radiation.

A number of intriguing behaviors were also found, including the crossing of curves in the time delay plots, and the angular deflection of geodesics during microlensing events pointing towards each other. It was also noted that the condition under which microlensing begins and ends in terms of the required distance of closest approach has an $O(1)$ correction that relates to the velocity of the appropriate segment of string. In the configurations investigated here, we found microlensing with impact parameters further from the string at the start and closer to it at the end than anticipated. Further investigations will be necessary to understand whether these observations hold more generally than the two situations analyzed here.

The basic goal of this program of work is to provide reliable information for the calculation of microlensing rates, as well as to provide accurate descriptions of microlensing events for observers.
Although the numerical examples presented involved unphysically large string tensions ($G \mu$ of $10^{-3}$ and $10^{-4}$ respectively) we identified how the phenomena scale with string tension so the results may be applied to physical string tensions of interest.

There is much further work to be done.
Our first task will be to generalize our code to handle kinks and straight segments of string. While these features will likely have interesting microlensing signatures in their own right, physical strings are also highly likely to possess kinks. Our second task will be to evaluate the magnification profile during microlensing events. While conventional wisdom suggests that flux doubles during a microlensing event, this is unlikely to be the full story, and it will be interesting to see just how the magnification changes during a microlensing event.

Once the tools to investigate these effects are in hand, there are two distinct lines of inquiry. The first involves investigating individual configurations in detail, with the aim of understanding the identified unexplained behaviors and identifying interesting physical effects in cosmic string microlensing. This would be particularly useful in order to understand the behavior of the magnification of images during microlensing events. This would also involve investigating effects from different string configurations, as well as the effect of small scale string structure.

The second path involves calculating the microlensing signatures for a large number of configurations of strings and source/observer positions. The aim would be to identify how the magnification profile, duration and frequency of microlensing events scale with loop size, string tension, observer and source distance, and string and source velocity. An accurate understanding of the appropriate scaling relations would allow for more precise rate predictions to be made, which would in turn lead to more accurate bounds on cosmic string tensions, and possibly even the observation of cosmic strings in our galaxy.

Ever-more capable
optical surveys are being planned and carried out, and we are hopeful that microlensing will give us a window into high energy processes from the earliest period of the universe's history.

\acknowledgments

We thank Eanna Flanagan, Leo Stein and Henry Tye for useful conversations. We gratefully acknowledge the support of the John Templeton Foundation (Univ. of Chicago 37426-Cornell FP050136-B).

\bibliographystyle{apsrev}
\bibliography{stringbib}

\end{document}